\documentclass[lettersize,journal]{IEEEtran}
\usepackage{amsmath,amsfonts}
\usepackage{algorithmic}
\usepackage{algorithm}
\usepackage{array}
\usepackage[caption=false,font=normalsize,labelfont=sf,textfont=sf]{subfig}
\usepackage{textcomp}
\usepackage{stfloats}
\usepackage{url}
\usepackage{verbatim}
\usepackage{graphicx}
\usepackage{cite}
\usepackage{multirow} 
\usepackage{graphicx} 
\usepackage{array}    
\usepackage{enumitem} 

\usepackage{tabularx}  
\usepackage{booktabs} 
\usepackage{amssymb}
\usepackage{orcidlink}
\usepackage{xcolor}

\usepackage{hyperref}
\hypersetup{
	colorlinks=true,
	citecolor=blue,
	linkcolor=blue  
}
\let\originalcite\cite
\renewcommand{\cite}[1]{{\color{blue}\originalcite{#1}}}

\usepackage{enumitem}
\usepackage{tikz}

\usepackage{subfig}
\usepackage{times}  

\usepackage{mdframed}

\begin{document}

\title{A Survey of Physical-layer Authentication Enhanced by Emerging Spatial Domain Technologies}
\author{Yuhao Chen\orcidlink{0009-0000-5050-1688},
	Boxiang He\orcidlink{0000-0002-9235-1144},
	Junshan Luo\orcidlink{0000-0003-4151-5126},
	Shilian Wang\orcidlink{0000-0003-4132-8750},
	Yiyan Ma\orcidlink{0000-0001-8034-6025},\\ \textit{Member, IEEE},
	Hao Xu\orcidlink{0000-0001-9847-7904}, \textit{Senior Member, IEEE},
	Lei Yao \orcidlink{0009-0005-3172-1462},
	Jing Lei\orcidlink{0000-0002-5838-5826},
	\\Arumugam Nallanathan\orcidlink{0000-0001-8337-5884}, \textit{Fellow, IEEE},
	and Kai-Kit Wong\orcidlink{0000-0001-7521-0078}, \textit{Fellow, IEEE} 
	\thanks{Yuhao Chen, Boxiang He, Junshan Luo, Shilian Wang, Lei Yao, and Jing Lei are with the College of Electronic Science and Technology, National University of Defense Technology, Changsha 410003, China(email: cyh20220720@163.com; boxianghe1@bjtu.edu.cn; luojunshan10@nudt.edu.cn; wangsl@nudt.edu.cn; yaolei11103@163.com; leijing@nudt.edu.cn).}

	\thanks{Yiyan Ma is with the School of Electronic and Information Engineering, Beijing Jiaotong University, Beijing 100044, China(email: mayiyan@bjtu.edu.cn).}

	\thanks{Hao Xu is with the National Mobile Communications Research Laboratory, Southeast University, Nanjing 210096, China(email: hao.xu@seu.edu.cn).}
	
		
	\thanks{Arumugam Nallanathan is with the School of Electronic Engineering and Computer Science, Queen Mary University of London, E1 4NS London, U.K.(e-mail: a.nallanathan@qmul.ac.uk).}
		
	\thanks{Kai-Kit Wong is with the Department of Electronic and Electrical Engineering, University College London, Torrington Place, WC1E 7JE, U.K., and also with the Department of Electronic Engineering, Kyung Hee University, Yongin-si, Seoul 17104, South Korea(e-mail: kai-kit.wong@ucl.ac.uk).}

}


\maketitle

\begin{abstract}
	This article surveys spatial-domain-enhanced Physical-layer Authentication (PLA), with Dual-polarized Antennas (DPA), Massive Multiple-Input Multiple-Output (MIMO), and Reconfigurable Intelligent Surfaces (RIS) as the primary focus. With the rapid growth of wireless deployments, authentication mechanisms face stringent requirements for high security, low overhead, and low latency. PLA offers lightweight identity verification by exploiting physical-layer characteristics. However, the effectiveness of PLA critically depends on how physical observations are constructed and validated under wireless channels. Unlike existing surveys that mainly organize PLA by authentication modality, feature source, and evaluation metrics, this work emphasizes the connection between spatial-domain enhancement mechanisms, the resulting feature representation, and the authentication procedure. We review how DPA, Massive MIMO, and RIS reshape PLA feature representation, and we summarize newly introduced security threats along with representative defense strategies. Case studies further illustrate the practical impact, such as representative detection-probability trends across Signal-to-Noise Ratio regimes and quantitative comparisons among representative schemes. Finally, we outline promising future opportunities enabled by Dynamic Metasurface Antennas, Extra-large MIMO, and spatial configuration with artificial intelligence.
\end{abstract}


\section{Introduction}
\label{sec1}
\subsection{Motivation}

\IEEEPARstart{T}{his} article is a survey that systematically reviews spatial-domain-enhanced Physical-layer Authentication (PLA) and primarily focuses on PLA schemes assisted by Dual-polarized Antennas (DPA), Massive Multiple-Input Multiple-Output (MIMO), and Reconfigurable Intelligent Surfaces (RIS) during the period from 2020 to 2025. 

With the rapid development of wireless communication technologies and the large-scale deployment of connected devices, the security authentication mechanisms of wireless networks face unprecedented challenges. Traditional upper-layer authentication mechanisms via cryptographic algorithms face several difficulties. Their security depends on computational assumptions that could be broken, and they are vulnerable to replay attacks. Additionally, they involve high communication and computational overheads, as well as complex key management \cite{Z25XIE,refC0330}. PLA is a promising auxiliary authentication approach that provides lightweight, low-latency, and secure identity verification solutions by utilizing wireless channel characteristics, device fingerprints, and other physical signals \cite{TIFS01}. The academic community has conducted extensive research and systematic surveys on PLA technologies. As shown in Table \ref{tbl1}, existing surveys classify PLA from different perspectives. Specifically, Xie et al. categorize PLA schemes into active and passive types \cite{Z25XIE}. Illi et al. further classify them as keyless PLA and key-based PLA \cite{refC0301}. Zhang et al. propose a Distributed Physical-layer Authentication (DPLA) framework and demonstrate that voting-based collaborative authentication significantly outperforms centralized schemes \cite{refQ0302}. Yang et al. conduct a systematic review of PLA techniques in backscatter communication (BC) \cite{refQ0401}. Lai et al. \cite{refC0330} are the first to analyze various PLA schemes from a signal processing perspective, categorizing existing schemes by signal processing domain and thoroughly investigating the performance of model-based and machine-learning-based authentication strategies across different processing orders.

\begin{table*}[tp]
	\caption{A comparative analysis of the contributions made by existing survey papers on PLA, with key classification frameworks.\label{tbl1}}
	\centering
	\renewcommand{\arraystretch}{1.45}
	\begin{tabular}{|c|c|m{7cm}|m{3.5cm}|}
		\hline
		\textbf{Authors} & \textbf{Year} & \textbf{Key Focus} & \textbf{Classification} \\
		\hline
		Xie et al.\cite{Z25XIE} & 2021 & Categorizes and overviews PLA techniques, and summarizes performance and application scenarios. & $\bullet$  Passive \newline$\bullet$ Active \\
		\hline
		Lai et al.\cite{refC0330} & 2025 & Systematically categorizes approaches with focus on signal processing methodologies in PLA. & $\bullet$ Model-driven \newline$\bullet$ Data-driven \\
		\hline
		Illi et al.\cite{refC0301} & 2024 & Categorizes PLA technologies into two main groups with distinct operational characteristics. &$\bullet$ Keyless \newline$\bullet$ Key-based\\
		\hline
		Zhang et al.\cite{refQ0302} & 2024 & Proposes a distributed PLA framework and verifies its superiority over alternative approaches. & $\bullet$ Distributed \newline$\bullet$ Centralized \\
		\hline
		Yang et al.\cite{refQ0401} & 2024 & Reviews PLA techniques in BC systems and proposes a comprehensive security threat model. &$\bullet$ BC-specific \\
		\hline
	\end{tabular}
\end{table*}

\begin{figure*}[t]
	\centering
	\includegraphics[width=0.8\linewidth]{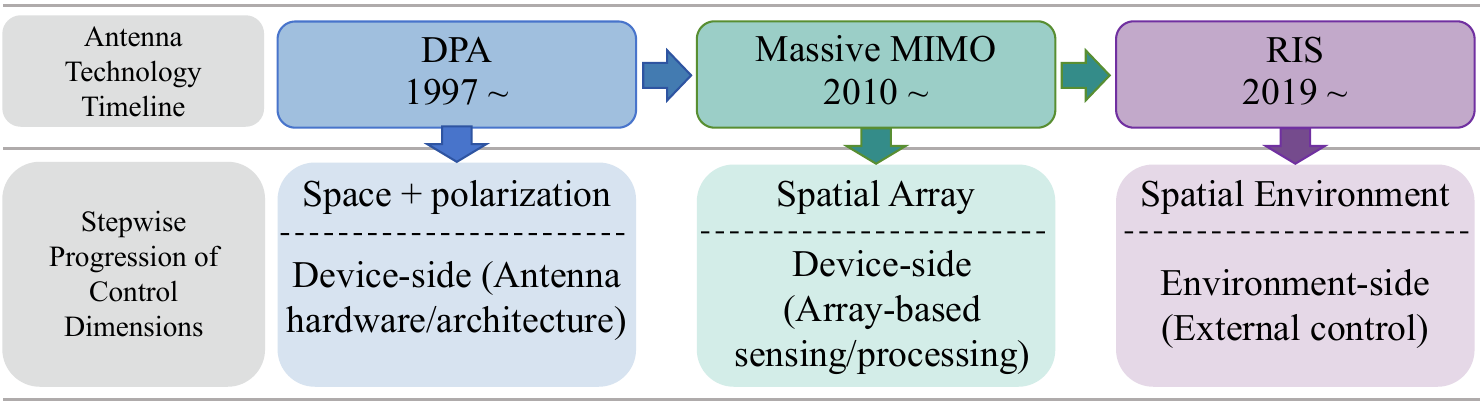}
	\caption{The organizing logic of the survey.}
	\label{fig:survey_logic}
\end{figure*}

The organizing principle that governs the structure of this survey is illustrated in Fig.~\ref{fig:survey_logic}. The survey proceeds according to the evolution of the spatial control position in the PLA. The discussion starts from device-side antenna hardware DPA in the space-polarization domain, advances to device-side array-based sensing and processing with massive MIMO in the spatial array domain, and ultimately reaches environment-side programmable reconfiguration through RIS in the spatial environment domain. This progression reflects a systematic expansion of the spatial degrees of freedom available to PLA systems and loosely follows the chronological maturity of the technologies, with DPA being the earliest and most established, massive MIMO representing a commercially mature and widely deployed technology, and RIS being an emerging technology that is actively investigated for future wireless systems. Following this evolution, this article sequentially focuses on the spatial domain enhancement effects of these three representative technologies on PLA. Emerging technologies such as Dynamic Metasurface Antennas (DMA), Hybrid RIS (H-RIS), Extra-large MIMO (XL-MIMO), and Fluid Antenna Systems (FAS) are discussed in Section \ref{sec6} as extensions or alternative developments within their respective categories, which highlights promising directions for future research as their PLA-specific studies continue to grow.

Across these works, a common thread is that PLA is typically organized along core technical elements such as authentication modality, feature sources, and performance evaluation metrics, and security discussion is generally scoped within the corresponding PLA pipeline. However, a unified framework has yet to be formed that systematically connects spatial-domain enhancement mechanisms to feature construction and authentication procedures. The effectiveness of PLA also depends on the ability of physical-layer observations, shaped by wireless channels and device hardware impairments, to capture the uniqueness of legitimate transmitters and links. The availability and richness of physical observations directly affect the efficacy of authentication in distinguishing between legitimate and adversarial users. This ultimately determines both detection performance and vulnerability to attacks.

With the emergence and maturation of spatial-domain enhancement technologies, the observable physical dimensions in PLA have been significantly expanded \cite{TII01}. We categorize spatial-domain enhancement technologies according to their primary contribution mechanisms and control positions in PLA, yielding three representative categories:

\begin{itemize}
	\item \textbf{Device-side Polarization Domain Expansion:}  Exploiting orthogonal polarization states provides additional authentication dimensions independent of spatial information. DPA has been extensively studied in PLA research, representing a well-established technology with a substantial body of literature demonstrating its effectiveness in authentication. Polarization-dependent authentication features can be extracted using this technology.
	
	\item \textbf{Device-side Array-based Spatial Resolution Enhancement:}  Antenna array design enhances PLA performance by increasing spatial resolution and observability. Massive MIMO has been studied for PLA applications and represents a well-established approach deployed in 5G systems. It exploits large-scale antenna arrays operating in the far-field regime with stationary channel characteristics to provide fine-grained spatial information for authentication.
	
	\item \textbf{Environment-side Programmable Channel Reshaping:}  Spatial-domain technologies in this category enhance PLA performance by modifying channels. Specifically, RIS exemplifies this approach by dynamically adjusting the phase shifts of reflecting elements to reshape wireless channels in a controllable manner.
\end{itemize}

\begin{figure*}[t]  
	\centering  
	\includegraphics[width=0.9\textwidth]{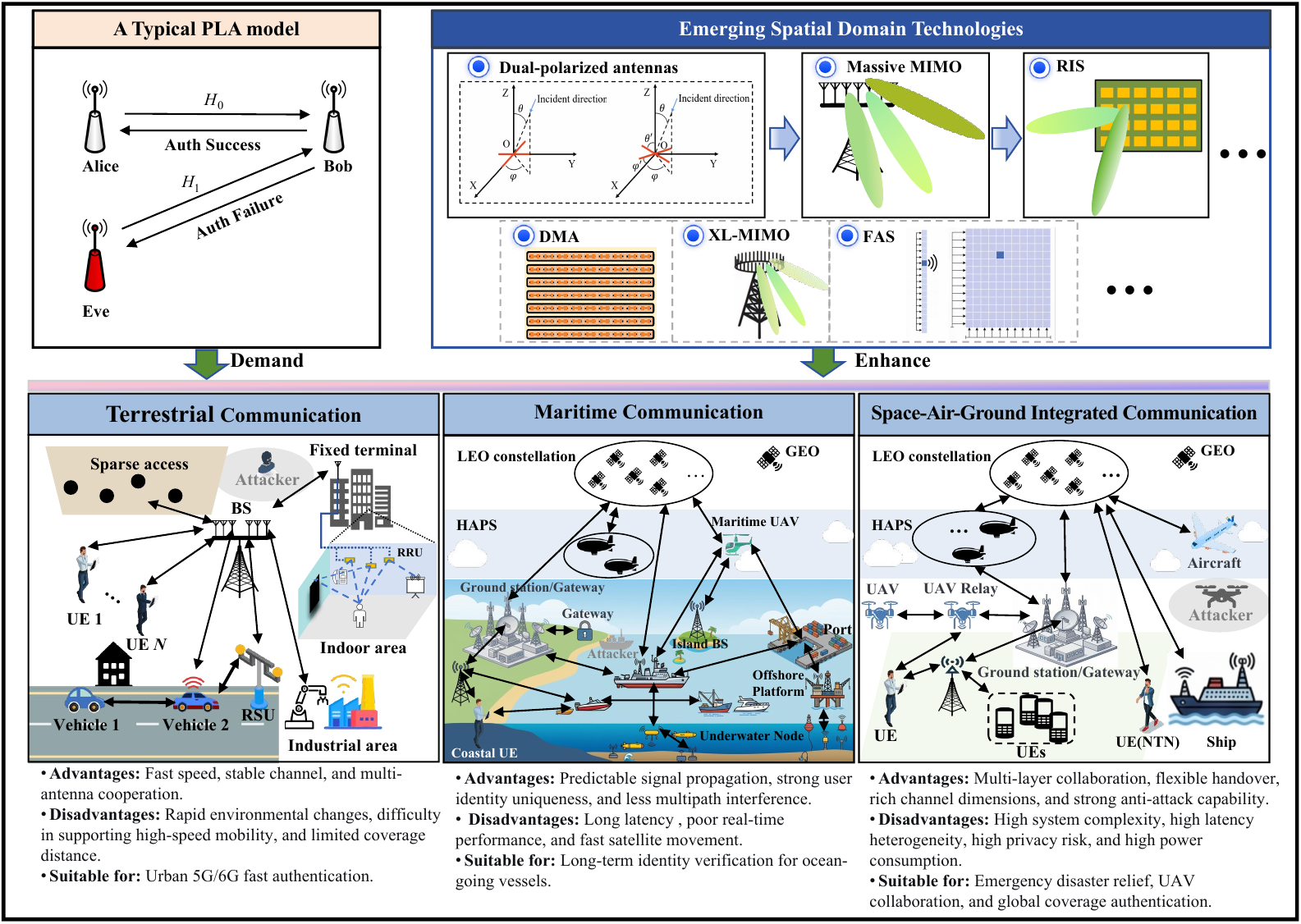}  
	\caption{PLA demand from diverse communication scenarios and enhancement via emerging spatial domain technologies.}
	\label{Framework_1}  
\end{figure*}

Fig.~\ref{Framework_1} provides a structured overview of this research landscape. The upper left panel summarizes a typical PLA hypothesis-testing framework with a legitimate transmitter-receiver pair and an adversary. The upper-right panel organizes the spatial-domain technologies targeted in this survey by their enhancement mechanisms, with DPA, massive MIMO, and RIS representing the three most extensively studied technologies. DMA, XL-MIMO, and FAS represent emerging technologies within or extending these categories and are discussed in future research directions. The bottom portion illustrates diverse communication scenarios across terrestrial, maritime, and space-air-ground integrated systems, highlighting the PLA requirements in different environments. Importantly, while Fig.~\ref{Framework_1} outlines the conceptual linkage between PLA demands and spatial-domain enhancement opportunities, the detailed mechanisms by which these technologies enhance physical-layer feature representation, together with the security threats and defense strategies they introduce, are systematically analyzed in the subsequent sections. Motivated by the above research status and technological development trends, this article provides a systematic survey of spatial-domain-enhanced PLA, aiming to offer a structured reference and identify key research avenues for developing secure authentication systems.

\begin{figure*}[t]  
	\centering  
	\includegraphics[width=0.95\textwidth]{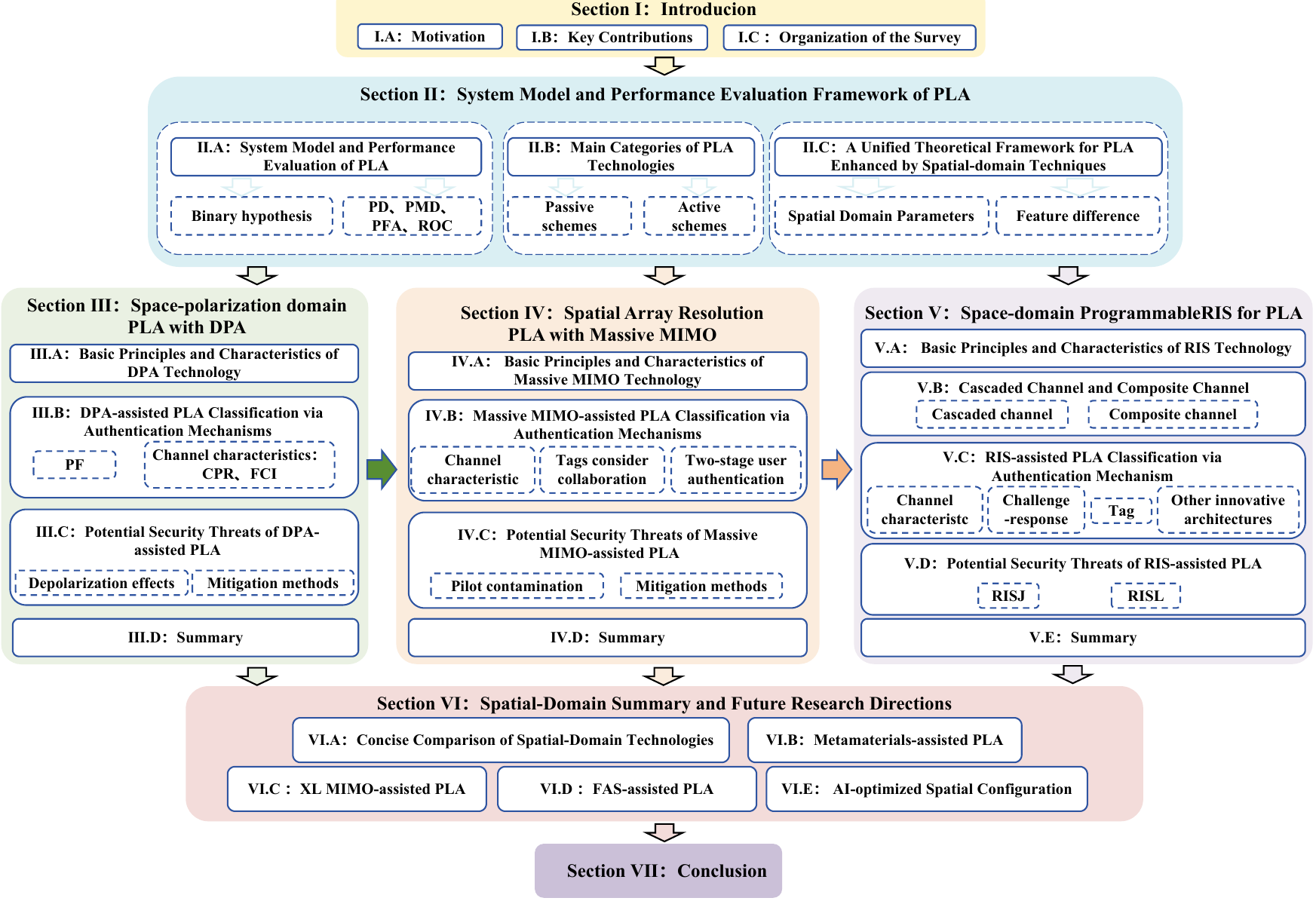} 
	\caption{The organization structure of the survey.}  
	\label{Organization}  
\end{figure*} 

\begin{table}[tbp]
	\caption{List of Abbreviations\label{tbl2}}
	\centering
	\renewcommand{\arraystretch}{1}
	\begin{tabular}{|c|c|}
		\hline
		\textbf{Abbreviations} & \textbf{Full Name} \\
		\hline
		3GPP    & 3rd Generation Partnership Project \\
		\hline
		AN      & Artificial Noise \\
		\hline
		AUDCP   & Additively Unique Decomposable Constellation Pair \\
		\hline
		BAN     & Burrows-Abadi-Needham \\
		\hline
		BER     & Bit Error Rate \\
		\hline
		CFO     & Carrier Frequency Offset \\
		\hline
		CDF     & Cumulative Distribution Function \\
		\hline
		CIR     & Channel Impulse Response \\
		\hline
		CNN     & Convolutional Neural Network \\
		\hline
        CPR     & Channel Polarization Response \\		
		\hline
		CSI     & Channel State Information \\
		\hline
		CVCA    & Complex-valued Classifiable Autoencoder \\
		\hline
		DBF     & Digital Beamforming \\
		\hline
		DGG     & Double Generalized Gamma \\
		\hline
		DMA     & Dynamic Metasurface Antenna \\
		\hline
		DPA     & Dual-polarized Antenna \\
		\hline
		FCI     & Full Channel Information \\
		\hline
		HBF     & Hybrid Beamforming \\
		\hline
		I/Q     & In-phase/Quadrature \\
		\hline
		IoT     & Internet of Things \\
		\hline
		LMMSE   & Linear Minimum-Mean-Square-Error \\
		\hline
		LOS     & Line-of-Sight \\
		\hline
		MeRFFI  & Metasurface RF-Fingerprinting Injection \\
		\hline
		MIMO    & Multiple-Input Multiple-Output \\
		\hline
		ML      & Maximum Likelihood \\
		\hline
		mmWave  & Millimeter-Wave \\
		\hline
		NLoS    & Non-Line-of-Sight \\
		\hline
		PDL     & Polarization Dependent Loss \\
		\hline
		PDF     & Probability Density Function \\
		\hline
		PD      & Probability of Detection \\
		\hline
		PFA     & Probability of False Alarm \\
	
		\hline
		PMD     & Probability of Missed Detection \\
		\hline
		PLA     & Physical Layer Authentication \\
		\hline
		PRR     & Phase Response Ratio \\
		\hline
		ROC     & Receiver Operating Characteristic \\
		\hline
		RIS     & Reconfigurable Intelligent Surface \\
		\hline
		RISJ    & RIS Jamming \\
		\hline
		RISL    & RIS Leakage \\
		\hline
		RSS     & Received Signal Strength \\
		\hline
		SEI     & Signal Emitter Identification \\
		\hline
		SER     & Symbol Error Rate \\
		\hline
		SIMO    & Single-Input Multiple-Output \\
		\hline
		SNR     & Signal-to-Noise Ratio \\
		\hline
		SRS     & Sounding Reference Signal \\
		\hline
		TCI     & Transmission Configuration Indication \\
		\hline
		UAV     & Unmanned Aerial Vehicles  \\
		\hline
	    AAV     & Autonomous Aerial Vehicles \\
		\hline
		VLC     & Visible Light Communication \\
		\hline
		VR      & Visibility Region \\
		\hline
		XL-MIMO & Extra-Large MIMO \\
		\hline
	\end{tabular}
\end{table}

\subsection{Key Contributions}
As these emerging technologies reshape the landscape of PLA research, new opportunities and challenges arise. To this end, this article presents a systematic and comprehensive survey of spatial-domain-enhanced PLA. The key contributions are summarized as follows:

\begin{enumerate}
	\item We review the foundational theory of PLA, encompassing core concepts, technical classifications, and performance evaluation metrics. For the first time, we systematically unify multiple spatial-domain technologies, including DPA, massive MIMO, and RIS, within the PLA framework for analysis. This effort fills a gap in existing surveys regarding the spatial domain.
	
	\item This survey provides a detailed analysis of the applications of three spatial-domain technologies in PLA. For DPA-assisted authentication, we elaborate on mechanisms that rely on polarization fingerprints and dual-polarized channel characteristics. For massive MIMO, we focus on authentication schemes according to channel characteristics and tags, as well as the impacts of precoding design and hardware impairments. For RIS-assisted authentication, we systematically categorize four types of authentication methods using channel characteristics, challenge-response, tag mechanisms, and innovative architectures. By revealing how these technologies enhance authentication performance through increased channel complexity and uniqueness, we provide technical references for practical deployment.
	
	\item Our analysis thoroughly explores the security challenges faced by spatial-domain technologies in PLA applications. For DPA, we analyze the impact of depolarization effects on authentication performance. For massive MIMO, we focus on the threats posed by pilot contamination attacks and their mitigation strategies. For RIS, we analyze RIS Jamming (RISJ) and RIS Leakage (RISL) attacks, and their corresponding defense mechanisms. Finally, we organize the applicable scenarios and limitations of existing defense strategies.
	
	\item We explore emerging spatial-domain technologies such as DMA, XL-MIMO, and FAS for PLA, analyze their underlying principles, representative authentication schemes, and technical challenges, and discuss spatial configuration with artificial intelligence (AI) as a key approach to enhance authentication performance.
\end{enumerate}

\subsection{Organization of the Survey}
The arrangement of the subsequent sections is as follows: Section \ref{sec2} begins with an exploration of the fundamentals of PLA, encompassing its core concepts, primary classifications, and performance evaluation metrics, followed by a unified theoretical framework for spatial-domain-enhanced PLA. This groundwork establishes a theoretical foundation for the subsequent technical discussions. Section \ref{sec3}, the discussion centers on the role of DPA technology in PLA. This section introduces the fundamental principles and design structure of DPA, and subsequently explains authentication methods via polarization fingerprints and dual-polarized channel features. Section \ref{sec4} presents the implementation of MIMO technology in PLA, covering system architecture, authentication methods using channel characteristics and tags, precoding design, and the impact of hardware impairments, along with potential mitigations. This section analyzes the challenges of pilot contamination. Finally, we analyze relevant security performance metrics and challenges. Section \ref{sec5} focuses on RIS-assisted PLA, providing a description of the RIS hardware architecture, common channel models, including cascaded and composite channels, and categorizing authentication mechanisms using channel characteristics, challenge-response methods, tag utilization, and innovative architectures. This section also analyzes common security threats faced by RIS and the corresponding defense strategies. Section \ref{sec6} provides a concise comparative summary of the three spatial-domain technologies within a unified framework, explores prospective technologies including DMA, XL-MIMO, and FAS as promising future research directions, and further discusses an AI-optimized spatial configuration that leverages deep reinforcement learning and generative models to maximize authentication feature separability. Finally, Section \ref{sec7} provides a summary of the core content and contributions of the survey. The overall structure of this survey is illustrated in Fig. \ref{Organization}. All abbreviations used throughout the article and their corresponding full terms are summarized in Table \ref{tbl2}.

\section{System Model and Performance Evaluation Framework of PLA}
\label{sec2}
In this section, we introduce the performance evaluation framework, main classifications, and inherent characteristics of PLA technologies. Specifically, we further establish a unified theoretical framework for spatial-domain-enhanced PLA.

\subsection{System Model and Performance Evaluation of PLA}
As shown in the three-node model within Fig. \ref{Framework_1}, a typical architecture of a PLA system comprises a legitimate transmitter (Alice), a legitimate receiver (Bob) responsible for authentication, and a potential adversary (Eve). The core task for Bob is to distinguish Alice's signals from those spoofed by Eve. This decision is fundamentally a binary hypothesis testing problem, which can be formalized as
\begin{equation}  
	\label{deqn_ex1a_1}  
	\begin{cases}   
		H_0: \text{The signal is legal} , \\
		H_1: \text{The signal is illegal},  
	\end{cases}  
\end{equation}  
where the performance of the binary hypothesis test is quantified by several standard metrics as follows:
\begin{itemize}
	\item Probability of Detection (PD): Correctly detecting $H_1$ when $H_1$ is true.
	\item Probability of False Alarm (PFA): Incorrectly accepting $H_1$ when $H_0$ is true.
	\item Probability of Missed Detection (PMD): Failing to detect $H_1$ when $H_1$ is true.
	\item Receiver Operating Characteristic (ROC): The trade-off curve between PD and PFA.
\end{itemize}
where the above four metrics are standard in detection theory \cite{Z25XIE, refQ0302}. Additionally, a security-oriented metric is defined as:
\begin{itemize}
	\item Probability of Security Authentication (PSA) \cite{XIE_PSA}: Defined as $\max\{P_{D,\text{Bob}} - P_{D,\text{Eve}}, 0\}$, where $P_{D,\text{Eve}}$ corresponds to the PD above, and $P_{D,\text{Bob}}$ denotes the probability of correctly accepting the legitimate transmitter.
\end{itemize}

\subsection{Main Categories of PLA Technologies}
According to \cite{Z25XIE}, PLA can be categorized into two primary classes, namely \emph{passive} schemes and \emph{active} schemes. The key difference is whether the transmitted message is actively modified. In particular, passive authentication schemes verify the transmitter identity by exploiting physical characteristics of the received signal~\cite{ref35,refCFR1,refMuti4}. 
Unlike passive schemes, active schemes authenticate the transmitter by embedding an authentication tag generated from a shared key into the transmitted waveform~\cite{refMuti1,refMuti2}. Since this tag-based verification does not require an additional training phase, active schemes typically offer high robustness and low communication latency. However, reserving resources for tag embedding may reduce the power (or rate) available for the information-bearing payload, and the security of the tag generation mechanism is crucial \cite{refZ10}. Therefore, active schemes must balance security guarantees with communication performance.  Collectively, both passive and active PLA mechanisms can be viewed as lightweight authentication approaches, and they commonly share the following performance attributes:

\begin{itemize}
	\item Security guarantee: 
	Passive schemes typically derive security from physical-layer features that are hard to replicate,
	whereas active schemes rely on key-dependent authentication tags, the resulting security level
	may range from information-theoretic to computational, depending on the underlying assumptions.
	
	\item Low overhead and latency: 
	Authentication decisions are made directly at the physical layer without invoking complex upper-layer
	protocols, significantly reducing processing delay and computational complexity.
	
	\item High compatibility: 
	PLA is designed to complement, rather than replace, existing upper-layer authentication mechanisms,
	enabling the construction of two-factor authentication systems that enhance overall security.
\end{itemize}

\begin{figure}[tp]
	\centering
	\includegraphics[width=3.45in]{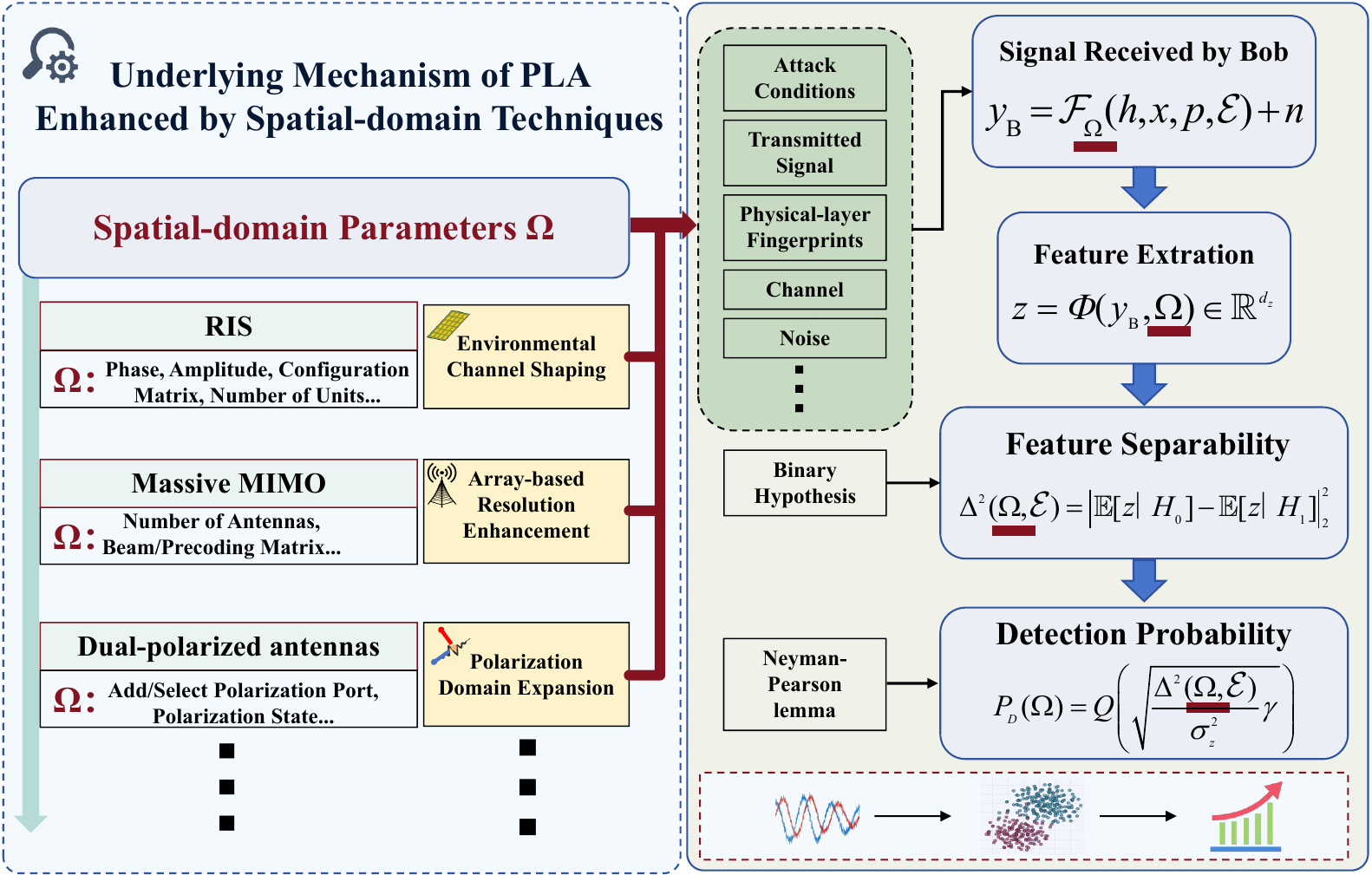}
	\caption{Schematic Diagram of the Underlying Mechanism of PLA Enhanced by Spatial-domain Techniques.}
	\label{Evaluation}
\end{figure}

\subsection{A Unified Theoretical Framework for PLA Enhanced by Spatial-domain Techniques}

As shown in Fig. \ref{Evaluation}, spatial-domain technology directly shapes the spatial characteristics of the received signal via parameter configuration in PLA. Specifically, the signal received by Bob under PLA can be expressed as
\begin{equation}
	y_{\text{B}} = F_\Omega(h, x, p, \mathcal{E}) + n,
	\label{eq:received_signal}
\end{equation}
where $y_{\text{B}} \in \mathbb{C}^{N_r}$ is Bob’s received signal vector; $F_\Omega(\cdot)$ is the system mapping function that combines the channel, transmitted signal, and physical fingerprint into the received signal; $\varOmega$ is the system configuration parameters, where for RIS they refer to phase, amplitude, configuration matrix, and number of units; for massive MIMO they refer to number of antennas and beam/precoding matrix; for DPA they refer to polarization port selection and polarization state. Furthermore, $h$ is the channel vector or matrix (legitimate or attack channel); $x$ is the pilot or data symbol sequence; $p$ is the physical fingerprint parameter vector; $\mathcal{E}$ is the environmental and attack conditions; $n \in \mathbb{C}^{N_r}$ is the noise and unmodeled disturbance vector, typically assumed to be complex Gaussian noise, $n \sim \mathcal{CN}(0, \sigma^2 I)$, where $\sigma^2$ is the noise power.

The authenticator maps $y_{\text{B}}$ to an authentication feature vector as~\cite{refC0330}
\begin{equation}
	z = \varPhi(y_{\text{B}}, \varOmega) \in \mathbb{R}^{d_z},
	\label{eq:feature_extraction}
\end{equation}
where $\varPhi(\cdot)$ is the feature extraction function jointly determined by the system, and $d_z$ is the feature dimension.

Following the binary hypothesis testing framework in \eqref{deqn_ex1a_1}, the separability is given by~\cite{5313937}
\begin{equation}
	\Delta^2(\varOmega, E) = \left| \mathbb{E}[z \mid H_0] - \mathbb{E}[z \mid H_1] \right|_2^2,
	\label{eq:separability}
\end{equation}
where $\mathbb{E}[z \mid H_0]$ is the expected feature vector under the legitimate user condition, $\mathbb{E}[z \mid H_1]$ is the expected feature vector under the attacker condition, and $|\cdot|_2$ is the Euclidean norm. This metric directly reflects the mean separation between the two feature classes under the $\Omega$ configuration.

According to signal detection theory~\cite{8108524,Kay1998}, under Gaussian assumptions, the detection probability can be expressed by the Neyman-Pearson, i,e.
\begin{equation}
	P_D(\varOmega) = Q\left(\sqrt{\frac{\Delta^2(\varOmega, \mathcal{E})}{\sigma_z^2} \gamma}\right),
\end{equation}
where $\sigma_z^2$ is the equivalent noise variance after feature extraction, i.e., the intensity of the original noise $n$ projected through the feature extraction function $\Phi(\cdot)$; $\gamma$ is the Signal-to-noise Ratio (SNR) of the received signal relative to noise; $Q(\cdot)$ is the Q-function, which is defined as~\cite{Kay1998}
\begin{equation}
	Q(x) = \frac{1}{\sqrt{2\pi}} \int_x^{\infty} e^{-t^2/2}  dt.
	\label{eq:q_function}
\end{equation}

\textbf{\textit{Lessons Learned:}} We first unify the system configuration parameters $\varOmega$ for spatial-domain techniques, including RIS, massive MIMO, and DPA, into the signal and feature extraction model for PLA. The definition of a feature is the separability between the legitimate user and the attacker across different configurations, and the relationships among PD, $\varOmega$, and SNR are presented. For PLA, by optimizing $var\Omega$, spatial-domain techniques can enlarge the feature difference between the legitimate user and the attacker, thereby improving the authentication detection probability. However, they employ essentially different enhancement approaches and underlying mechanisms. Subsequent sections will provide a survey of the applications of these three primary techniques in PLA.

\begin{figure}[t]
	\centering
	\includegraphics[width=0.85\linewidth]{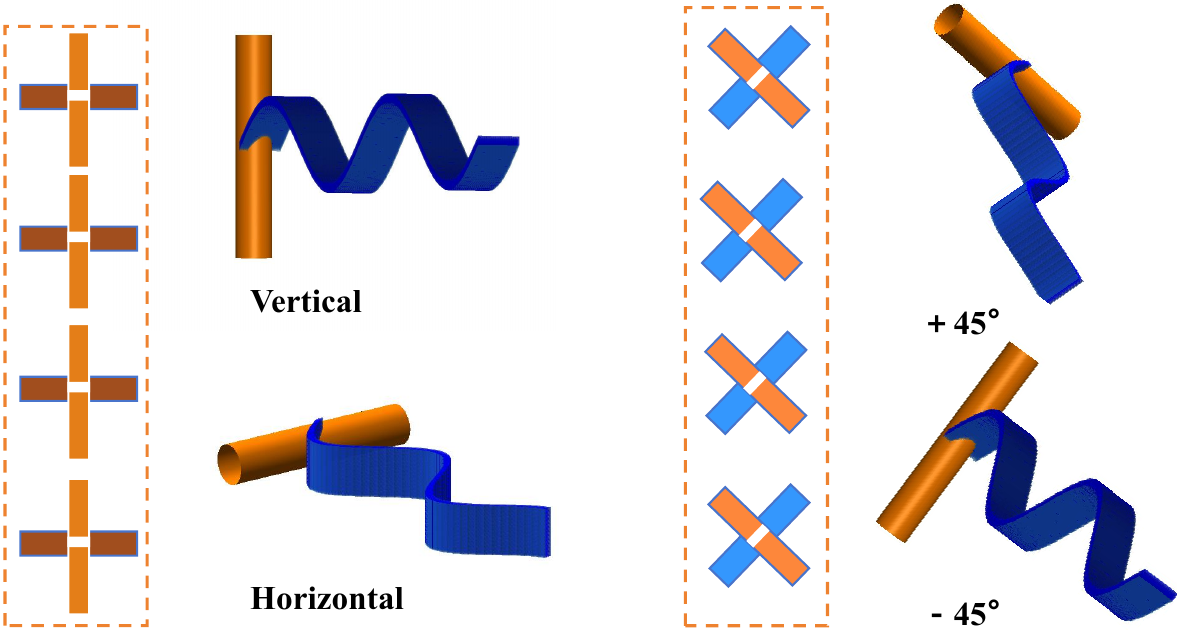}
	\caption{Diagram of Vertical/Horizontal Dual-polarized and $\pm45^\circ$ DPA.}
	\label{fig:dual-polarized-antennas}
\end{figure}

\section{Space-polarization domain PLA with DPA}
\label{sec3}
This section focuses on how DPA technologies support PLA. First, the key dual-polarization properties relevant to PLA are summarized. Two authentication mechanisms assisted by DPA are then introduced. Finally, security threats arising from depolarization effects in DPA systems are analyzed, and possible mitigation strategies are outlined.

\subsection{Basic Principles and Characteristics of DPA Technology}
DPA exploits the vector nature of polarization. In a dual-polarized configuration, an antenna can transmit and/or receive in two orthogonal polarization states. These states include vertical and horizontal polarizations as well as plus and minus $45^\circ$ diagonal polarizations, as illustrated in Fig. \ref{fig:dual-polarized-antennas}. This is achieved by integrating two mutually orthogonal radiating units~\cite{J05-40}. This enables the receiver to capture polarization-domain features that are difficult to replicate without reproducing the polarization-dependent wireless propagation and hardware-induced deviations.

From the PLA perspective, DPA offers two advantages. First, orthogonal polarization channels share the same frequency resource, which improves system capacity and supports polarization-domain multiplexing~\cite{J002,J005-3}. Second, polarization diversity and dual-polarized propagation introduce polarization-dependent fading and depolarization effects that can be leveraged as authentication evidence~\cite{J005-1,J005-4,J005-5,J005-6,J005-8,J005-10}. In particular, different transmitters and propagation environments lead to distinguishable polarization states and polarization-dependent channel responses, which can strengthen anti-spoofing performance.

Compared with single-polarized designs, a conventional DPA implementation typically employs two orthogonally polarized radiating elements. Common examples include microstrip patches~\cite{J005-42,J005-44}, magneto-electric dipoles~\cite{J005-25,J005-35}, or waveguide slit antennas~\cite{J005-37,J005-77}. To ensure adequate polarization isolation, designers often use independent feed networks and layout or common-ground strategies. These measures help reduce inter-polarization interference~\cite{J002}. For PLA, the dual-polarization interface can reliably deliver polarization-separated observations whose polarization-dependent distortions and depolarization-induced variations reflect the physical environment and transmitter-dependent effects, thereby providing rich polarization-domain features.

\subsection{DPA-assisted PLA Classification via Authentication Mechanisms}
This subsection provides a systematic classification of existing DPA-assisted PLA techniques based on their authentication mechanisms. We analyze and group methods into two main streams: polarization fingerprint (PF)-based approaches and channel-characteristics-based approaches. The classification is structured by the main authentication decision criterion. PF-based schemes mainly use polarization-fingerprint template matching. Device identity is verified using normalized distance metrics between estimated and reference templates. Channel-characteristics-based schemes instead use polarization-related channel responses estimated from received dual-polarized signals and apply hypothesis testing frameworks for PLA.

As a common abstraction, a dual-polarized DPA receiver forms polarization-separated observations at multiple frequency samples, which are then mapped to polarization-domain features and embedded into a binary-hypothesis testing framework.

\subsubsection{Utilizing DPA-PF for PLA}
Radio Frequency Fingerprint (RFF) identification offers advantages such as low complexity and high security. However, its discrimination capability has been gradually declining. Current RFF research primarily focuses on extracting physical-layer features from the time-frequency domain of RF signals. The polarization domain has been underutilized. This results in an incomplete characterization of these features. Novelty, Xu et al. proposed the concept of PF, which extracts physical-layer features representing device identity from the polarization domain of RF signals~\cite{J01,J02,J03}. Compared to conventional RFF, the primary advantage of PF lies in its vectorial nature, which significantly enhances fingerprint discrimination capability. This subsection will elaborate on the application of polarization fingerprints derived from different foundational frameworks in DPA-assisted PLA.

\begin{itemize}
	\item \textbf{Frequency-dependent polarization state curves:}
	Xu et al. are the first to introduce and rigorously demonstrate the existence of a polarization fingerprint inherent in wireless signals, thereby providing a novel physical-layer basis for wireless device authentication~\cite{J01}. Through a comprehensive mathematical modeling approach and empirical validation, the authors elucidate the distinctiveness of the polarization fingerprint, with particular attention to its frequency-dependent variation and its uniqueness arising from antenna structure and hardware imperfections. The polarization fingerprint is modeled as a function of ideal design parameters perturbed by hardware-induced imperfections. This model reveals three fundamental characteristics:
	\begin{enumerate}[label=\textcircled{\small\arabic*}]
		\item Group feature captures the structural information of the antenna, resulting in markedly distinguishable polarization fingerprint patterns among different antenna models.
		\item Individual feature reflects the fact that even antennas of the same model exhibit distinct polarization fingerprints due to manufacturing variability and individual discrepancies.
		\item Directionality emerges from the vectorial nature of polarization, whereby signals arriving from different incident angles induce deformation in the polarization fingerprint, thereby amplifying inter-device fingerprint diversity.
	\end{enumerate}
	
	\hspace*{1em}The directionality property can be described by two cases: when orthogonal dipole antennas are aligned with the coordinate axes, the received polarization is a function of the incident signal weighted by the cosine of the elevation angle; when the antennas are deflected, the polarization additionally depends on both the elevation and azimuth offset angles through a more complex expression.
	
	Moreover, the authors propose a polarization-fingerprint-based authentication framework for wireless IoT devices that comprises two principal stages: template database construction and identity verification. The authenticator collects polarization fingerprints from individual devices to build a reference template library. Upon receiving an authentication request, the system extracts the PF from the received signal and performs Euclidean distance-based matching against the stored templates to verify the identity of the device. Compared to conventional RFF authentication methods, which suffer from degraded accuracy at low SNR and limited sampling rates, thereby hindering reliable authorization~\cite{J01-10,J01-22}, the PF authentication approach effectively mitigates these challenges, achieving a simpler implementation and superior authentication accuracy.
	
	\item \textbf{Distinct differentiation between frequency and spatial characteristics:}
	In subsequent research, Xu et al. further propose leveraging the spatial characteristics of the PF for Specific Emitter Identification (SEI)~\cite{J03}. They note that traditional RFFs exhibit diminished hardware-induced variance due to advancements in manufacturing processes~\cite{J03-1,J03-7,J03-8}, thereby limiting their discriminatory capabilities. In contrast, PF not only includes frequency characteristics derived from antenna hardware imperfections but also possesses unique spatial attributes that capture the vectorial nature of signal polarization~\cite{J03-9,J03-10}. These spatial attributes remain invariant despite improvements in manufacturing processes, thus retaining a high degree of distinctiveness. The mathematical model combines the frequency-dependent hardware response with a spatial response that depends on the signal's elevation and azimuth angles of incidence. This model enables the authors to propose a method that records the corresponding PF for emitter identification in static scenarios, where the emitter's position is fixed. In dynamic scenarios involving moving emitters, the system introduces a Spatial Polarization Compensation (SPC) technique. This technique corrects PF variations caused by changes in emitter position using pre-measured transformation vectors, thereby maintaining identification accuracy. Performance evaluations demonstrate that the spatial characteristics of PF outperform traditional RFF methods~\cite{J01-10,J01-22} under low SNR conditions, and that the SPC approach effectively mitigates recognition errors in dynamic environments.
	
	\item \textbf{Integrating frequency and spatial characteristics:}
	Building on the foundational work established in earlier studies~\cite{J01,J03}, Xu et al. further refine the PF construction model, specifically targeting practical scenarios such as LoRaWAN IoT systems~\cite{J02}. Referring to a classical corner-truncated microstrip patch antenna model~\cite{J02-30}, the authors formulate a comprehensive PF model that integrates hardware-induced frequency response with spatial polarization responses at both the transmitter and receiver. This model captures hardware imperfections that affect polarization states across frequencies, while the spatial characteristics reveal non-uniform directional distributions of polarization vectors at both the transmitter and the receiver. This significantly enhances the discriminability among devices of the same model and improves overall system security. To address the PF fragmentation caused by frequency hopping in LoRaWAN, the authors proposed dividing the PF into multiple frequency segments. They then employed an ensemble CNN to independently learn each segment, followed by weighted soft voting to fuse the segment predictions. This approach effectively improves recognition accuracy for fragmented PFs. Furthermore, the introduced PF tracking authentication mechanism mitigates PF variations caused by device mobility, ensuring continuous and robust identity verification in dynamic environments.
\end{itemize}

Motivated by the above PF construction, we further evaluate the performance of DPA-assisted PLA using PF matching under a polarization-fingerprint-based decision rule. Specifically, the detection probability is characterized, with the authentication decision made by comparing the estimated PF with stored reference templates. A total of $10000$ Monte Carlo trials are conducted for each simulation, and the target PFA is set to $0.01$ via quantile-based threshold calibration. The correlation coefficient between Alice's and Eve's channels is set to $0.6$, while the depolarization factor is set to $0.15$.

\begin{figure}[t]
	\centering
	\includegraphics[width=0.425\textwidth]{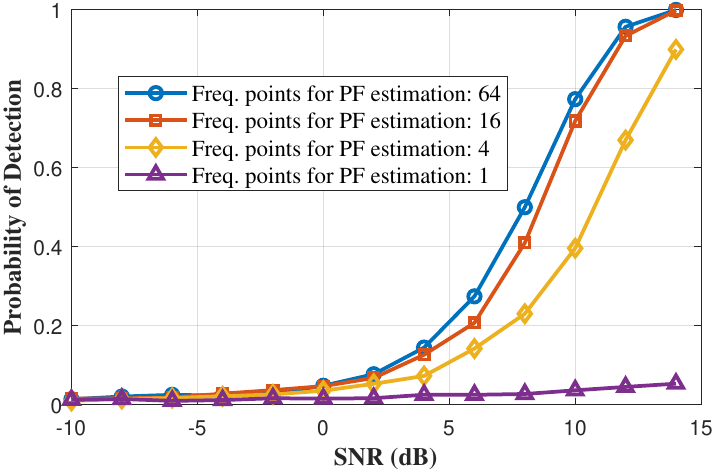}
	\caption{PD versus SNR for DPA-assisted PLA. Authentication is validated via PF matching using the normalized Euclidean distance metric.}
	\label{fig:case_study_dpol_pd_vs_snr}
\end{figure}

\begin{figure}[htbp]
	\centering
	\includegraphics[width=0.42\textwidth]{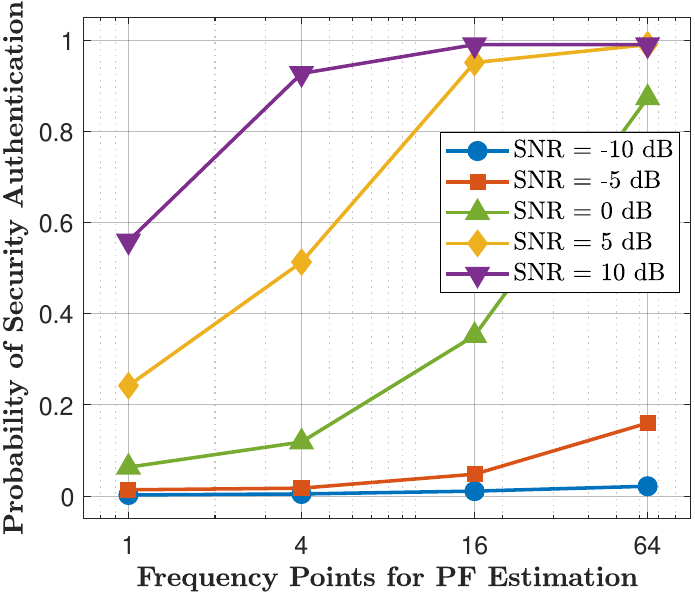}
	\caption{PSA of DPA-assisted PLA versus the number of frequency sampling points for polarization fingerprint estimation.}
	\label{fig:dpa_psa}
\end{figure}

Here, the authentication distance metric is the normalized Euclidean distance between the estimated PF and the reference template, which is given by
\begin{equation}
	D_{\text{PF}} = \left\|\frac{\hat{p}}{\|\hat{p}\|} - \frac{p_{\text{ref}}}{\|p_{\text{ref}}\|}\right\|_2^2,
	\label{eq:pf_distance_dual}
\end{equation}
where $\hat{p}$ denotes the estimated polarization fingerprint and $p_{\text{ref}}$ represents the stored reference template.

As shown in Fig.~\ref{fig:case_study_dpol_pd_vs_snr}, when the number of frequency sampling points increases from $1$ (corresponding to the baseline scalar polarization state matching scheme) to $64$, the detection probability improves from $0.32$ to $0.78$. This result highlights the crucial role of frequency sampling density in authentication reliability: the baseline scheme exploits only single-point polarization information, whereas multi-frequency sampling schemes leverage cross-band polarization variations to provide stronger discriminability, thereby enhancing authentication robustness. Moreover, the PFA control achieves a mean deviation of only $0.05\%$ from the target value of $0.01$. This indicates strong resilience to hardware impairments and depolarization-induced degradation, further supporting the practical effectiveness of DPA-based authentication.

Fig.~\ref{fig:dpa_psa} demonstrates how the PSA depends on the number of frequency sampling points at various SNR levels. As the number of sampling points increases, PSA rises consistently across all SNR regimes, reflecting that multi-frequency sampling enables more effective capture of polarization variations and enhances feature separability. For instance, at SNR = $0$~dB, the PSA increases from $0.0633$ for a single frequency point to $0.8733$ for $64$ frequency points.

\subsubsection{Utilizing dual-polarized channel characteristics for PLA}
Although the aforementioned work provides an in-depth discussion of the theory and application of PF, two recent studies~\cite{J05,J04}, conducted independently from the perspective of physical-layer dual-polarization channel characteristics, propose novel authentication methods that expand the dimensions of PLA techniques.

\begin{itemize}
	\item \textbf{Channel characteristics analysis via single polarization response:}
	 Wu et al. propose a novel PLA scheme using the Channel Polarization Response (CPR) in DPA communication systems, addressing critical limitations of existing PLA methods in highly dynamic scenarios, ultra-low SNR environments, and co-located adversarial attacks~\cite{J05}. The proposed scheme exploits the high sensitivity of CPR to signal polarization, enabling it to more effectively discriminate between distinct wireless channels than conventional Channel Frequency Response (CFR)-based methods~\cite{J05-2, J05-25, J05-30}. This capability is notably enhanced in rich scattering environments. By coherently stacking polarization state measurements within the channel coherence time, the scheme significantly improves the SNR, thereby enhancing CPR estimation accuracy and authentication reliability under extremely low SNR conditions. Furthermore, the continuous estimation characteristics of CPR allow adaptive adjustment of the authentication interval according to channel dynamics, making the scheme robust in high-mobility wireless environments. Additionally, leveraging hardware-induced polarization-state variations across devices enables discrimination of multiple transmitters at the same spatial location, effectively mitigating co-located attack threats. The proposed authentication framework comprises three principal stages.
	
	\begin{enumerate}[label=\textcircled{\small\arabic*}]
		\item \text{Coherent stacking of polarization states:} The received polarization state measurements within the channel coherence time are coherently accumulated. This accumulation effectively suppresses noise and scales the signal-to-noise ratio by a factor proportional to the number of stacked measurements. The stacked dual-polarized observation matrix is then modeled as the product of the transmitted polarization state matrix and the CPR vector, plus accumulated noise.
		
		\item \text{CPR estimation:} Leveraging the vertical and horizontal components of the dual-polarized signal along with their phase difference, the CPR is accurately estimated via a discrete Fourier transform in conjunction with inverse operations on the transmitted polarization states. The estimation process accounts for hardware impairments and noise, with a statistical model characterizing the resulting estimation errors.
		
		\item \text{Authentication via CPR:} A hypothesis testing framework is established utilizing a log-likelihood ratio test statistic that compares CPR estimates between consecutive authentication intervals to verify the identity of the transmitter. An optimal thresholding strategy achieves a balanced trade-off between false alarm probability and detection performance. Moreover, the scheme exploits the sensitivity of the CPR to polarization-state discrepancies induced by hardware imperfections across different transmitting devices, enabling the differentiation of multiple transmitters occupying the same spatial coordinates. This effectively addresses and mitigates the challenges posed by co-located attacks.
	\end{enumerate}
	
	\item \textbf{Joint characteristics of spatial fading and polarization fading in dual-polarized channels:}
	Unlike methods that rely solely on single-polarization responses, Wu et al. propose a novel PLA scheme that leverages Full Channel Information (FCI) in DPA communication systems~\cite{J04}. This approach transcends the limitations of traditional techniques, which exploit only channel spatial fading, by fully leveraging the joint spatial and polarization fading. Consequently, it enables a more comprehensive channel characterization and identity verification. When the receiver is configured with orthogonal DPA, the wireless channel is modeled as the sum over multipath components of spatial fading factors multiplied by polarization scattering matrices~\cite{J05-40}. Each spatial fading component captures path loss characteristics influenced by multipath propagation and Doppler effects, while each polarization fading component reflects polarization state variations induced by Polarization Mode Dispersion and Polarization Dependent Loss associated with scatterers.

	Notably, the scheme comprises the following essential steps:
	\begin{enumerate}[label=\textcircled{\small\arabic*}]
		\item \text{Independent estimation of FCI components:} By synchronizing and demodulating the received signals, pilot sequences are extracted to separately estimate the spatial fading and polarization fading. The spatial fading is obtained by synthesizing the vertical and horizontal polarization components at the receive antennas. Polarization fading is estimated by leveraging the transmit polarization states as ``pilot signals,'' with the polarization amplitude ratio and phase difference estimated via least-squares and modeled as Gaussian random variables.
		
		\item \text{Modeling temporal correlations:} The temporal evolution of FCI parameters among consecutive time slots, resulting from scatterer mobility, is characterized by a first-order autoregressive process. The temporal correlation coefficients for the spatial and polarization fading components effectively capture the time-variant nature of the channel.
		
		\item \text{Authentication decision via FCI:} Authentication statistics are constructed by evaluating the normalized Euclidean distance between FCI estimates over two consecutive time instants. The test statistic is decomposed into three components: one quantifying the spatial fading variation, one measuring the polarization amplitude ratio deviation, and one capturing the polarization phase difference discrepancy. Each component is normalized by its corresponding variance, which accounts for both temporal channel variations and noise. The overall statistic is used within a hypothesis-testing framework for sender identity verification. Simulation results demonstrate that the proposed scheme effectively exploits the uniqueness of polarization fading and the temporal correlation in spatial fading, outperforming traditional CFR-based methods~\cite{J04-10,J04-13} under low SNR and time-varying fading conditions.
	\end{enumerate}
	\end{itemize}

\begin{figure}[t]
	\centering
	\includegraphics[width=0.85\linewidth]{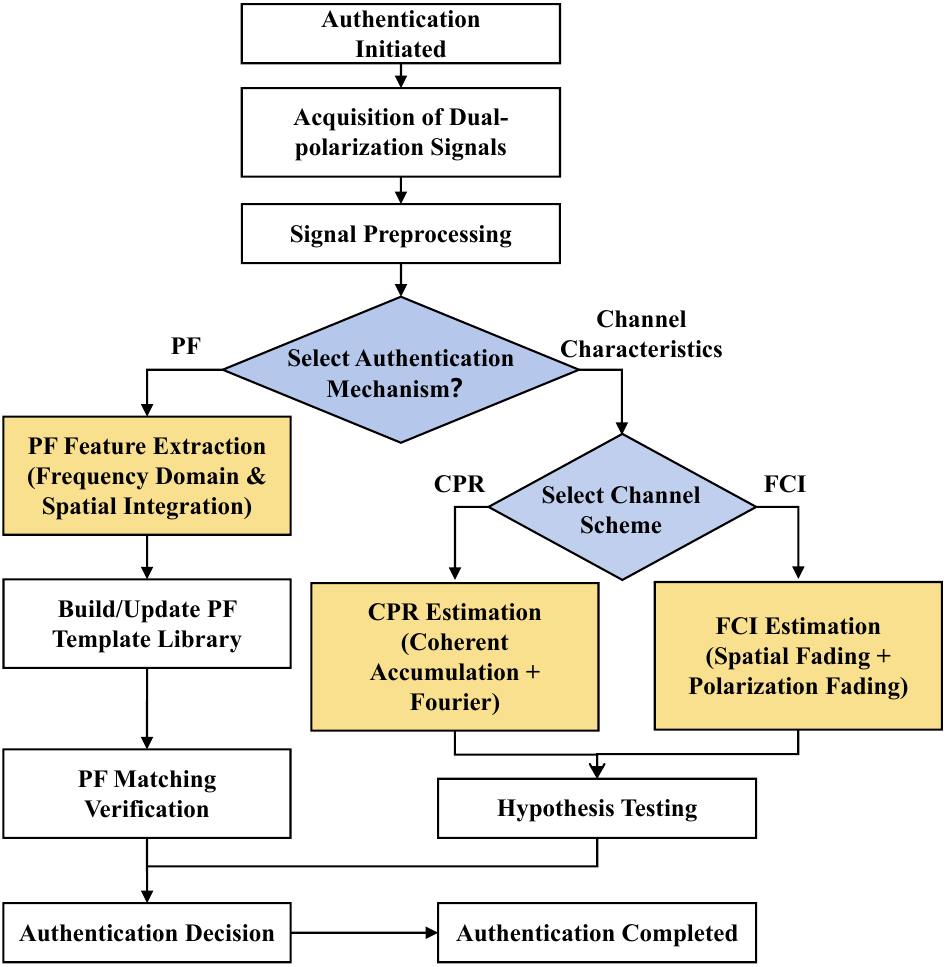}
	\caption{The two mainstream approaches for PLA assisted by DPA: PF-based and channel characteristics-based methods.}
	\label{Two_way_dual}
\end{figure}

In summary, the method proposed in~\cite{J05} utilizes the channel polarization characterization of single-polarization responses to improve authentication accuracy by leveraging the sensitivity of the polarization state. Conversely, the technique introduced in~\cite{J04} adopts a more comprehensive channel characterization, integrating full channel information that encompasses both spatial and polarization fading, thereby enhancing authentication performance. Drawing on the above discussion, DPA-assisted PLA can be summarized into two mainstream approaches: the PF-based method and the channel-feature-based method. Specifically, Fig.~\ref{Two_way_dual} presents the general flow of these two approaches.

\textbf{\textit{Lessons Learned:}} DPA-assisted PLA typically uses two main approaches. These include polarization-fingerprint-based authentication and channel-characteristics-based authentication. Polarization fingerprint methods use antenna hardware imperfections and frequency-dependent polarization states. Channel-characteristics methods combine spatial and polarization fading characteristics.

\subsection{Potential Security Threats of DPA-assisted PLA}
The security threats of DPA-assisted PLA technology primarily stem from signal quality degradation due to depolarization effects and increased hardware complexity.

\subsubsection{Security threats caused by depolarization effects}
The depolarization effect, which manifests as polarization-state confusion and perturbations during electromagnetic wave propagation, can substantially degrade the identification quality of polarized signals. It may induce crosstalk between vertically and horizontally polarized signals, leading to aliasing and attenuation of polarization states~\cite{J004-20}. Combined with polarization-related losses, this causes SNR deterioration and instability in polarization parameter extraction, thereby increasing the false identification rate. Furthermore, polarization feature degradation compromises eavesdropping-prevention capabilities~\cite{J004-40}, as signal obfuscation provides attackers with opportunities to forge or steal authentication information, elevating overall security risks.

\subsubsection{Mitigation methods for the impact of depolarization effects on DPA-assisted PLA}
Although a detailed discussion exceeds the scope of this article, two primary approaches have proven effective. First, high-precision polarization channel modeling and compensation algorithms address signal power cross-leakage by estimating and compensating for cross-polarization components~\cite{J004-2,J004-6}. Second, multi-dimensional joint modulation techniques introduce a polarization dimension into traditional amplitude-phase modulation, improving spectrum utilization and anti-interference capability while mitigating performance degradation caused by depolarization~\cite{J004}.

\begin{table*}[t]
	\caption{Summary of Antenna Types and Usage in Polarization Fingerprint Papers\label{tab:polarization_fingerprint_summary}}
	\centering
	\renewcommand{\arraystretch}{1.5}
	\begin{tabular}{|m{0.06\textwidth}|m{0.03\textwidth}|m{0.14\textwidth}|m{0.13\textwidth}|m{0.46\textwidth}|}
		\hline
		\textbf{Reference} & \textbf{Year} & \textbf{Tx Antenna} & \textbf{Rx Antenna} & \textbf{Usage Description} \\
		\hline
		\cite{J01} & 2022 & Circularly polarized patch antenna & Orthogonal DPA & Tx antenna defects produce unique and stable polarization fingerprints; Rx antenna captures frequency and direction-dependent polarization features for device authentication. \\
		\hline
		\cite{J03} & 2022 & Circularly polarized patch antenna & Orthogonal DPA & Tx hardware defects cause frequency-dependent polarization features; Rx antenna leverages spatial polarization to enhance device distinction. \\
		\hline
		\cite{J02} & 2023 & Circularly polarized patch antenna & Orthogonal DPA & Tx antenna manufacturing and structure induce unique frequency and spatial polarization features; Rx antenna extracts dual-polarized components; DL method handles fingerprint fragmentation. \\
		\hline
		\cite{J05} & 2023 & Vertical polarization antenna & $\pm 45^\circ$ DPA & Tx sends fixed polarization signals with subtle hardware variations; Rx captures multi-polarization information for channel polarization response estimation, enhancing authentication robustness. \\
		\hline
		\cite{J04} & 2023 & Orthogonal DPA & Orthogonal DPA & Tx and Rx antennas transmit and receive vertical and horizontal polarized signals, capturing polarization fading and spatial variations to improve channel information completeness and authentication accuracy. \\
		\hline
	\end{tabular}
\end{table*}

\subsection{Summary}  
We discuss DPA hardware architectures, the fundamental principles of polarization orthogonality, authentication mechanisms via polarization fingerprints and channel characteristics, and depolarization challenges. The key gain of DPA-assisted PLA is the introduction of polarization as a new discrimination dimension, which enables joint representation of device uniqueness and propagation differences. Depolarization weakens polarization distinguishability, induces feature confusion, and degrades authentication reliability. Future designs must balance enhancing polarization features and mitigating depolarization-induced degradation, with emphasis on improving polarization channel modeling and parameter compensation methods.



\section{Spatial Array Resolution PLA with Massive MIMO}
\label{sec4}
This section explores the application of massive MIMO technology in PLA. We first introduce the fundamental principles and technical characteristics of massive MIMO. Subsequently, we systematically categorize massive MIMO-assisted PLA mechanisms according to different authentication approaches: channel-characteristic methods, tag-assisted cooperative and non-cooperative schemes, and two-stage authentication frameworks. Finally, we discuss the security threats posed by pilot contamination in massive MIMO systems and examine corresponding mitigation strategies.

\subsection{Basic Principles and Characteristics of Massive MIMO Technology}
Massive MIMO is a crucial technology in wireless communications. The fundamental principle of MIMO enables concurrent signal processing by employing a significantly larger number of antennas on both the base station and user equipment \cite{refM001,refM001-2}. This technology relies on spatial multiplexing and beamforming, markedly enhancing the spectral and energy efficiency of the system by effectively utilizing the spatial resources of the wireless channel. The functionality of massive MIMO is underpinned by the following mechanisms.

\begin{itemize}
	\item Spatial multiplexing: By utilizing multiple antennas on the device, the spatial dimensions of the channel are exploited to simultaneously transmit independent data streams to multiple users \cite{refM001-7,refM002}.
	\item Beamforming: The base station directs signal energy toward specific users by adjusting the transmission weights of the antenna array, thereby minimizing interference and power loss \cite{refM001-7,refM002-13}.
	\item Accurate channel state information: The base station achieves precise signal processing by obtaining accurate channel state information (CSI) \cite{refM002-35,refM002-44}.
\end{itemize}

\begin{figure}[t]
	\centering
	\includegraphics[width=0.95\linewidth]{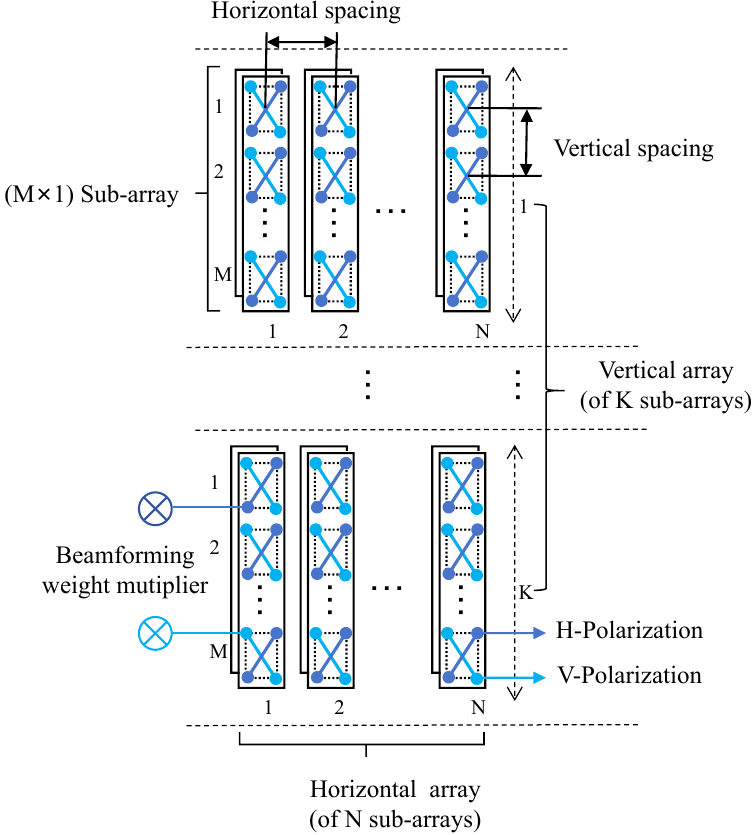}
	\caption{The architecture of a massive MIMO antenna array system.}
	\label{MASSIVE-MIMO-architecture}
\end{figure}

\subsubsection{Structure and characteristics of the Massive MIMO base station}
The design of a massive MIMO base station comprises three key components: dual-polarized microstrip antenna units, subarrays, and overall antenna arrays. While the antenna unit design is structurally similar in both the FR-1 (below 6 GHz) and FR-2 (millimeter-wave, above 24 GHz) categories, the shorter wavelength of FR-2 enables denser, more compact designs that are better suited for transmitting and receiving signals in higher frequency bands \cite{refM004-20}. The dual-polarized microstrip antenna unit can simultaneously transmit and receive orthogonal signals, thereby enhancing polarization diversity while maintaining a compact size through high cross-polarization isolation. The subarray consists of neighboring antenna units, which are typically utilized for digital beamforming (DBF) in FR-1, whereas in FR-2, it predominantly relies on phase-shift networks for analog or Hybrid Beamforming (HBF) to reduce hardware complexity and power consumption. Antenna arrays in massive MIMO systems are typically organized as two-dimensional uniform planar arrays (2D UPA). By integrating multiple subarrays, the system can adaptively adjust the beamforming direction to achieve directional signal transmission for specific users \cite{refM004}. Here, Fig.~\ref{MASSIVE-MIMO-architecture} presents the architecture of a massive MIMO antenna array.

\subsubsection{Antenna characteristics and factors influencing gain}
Antenna gain is a critical factor influencing the performance of massive MIMO, with determining factors including the number of subarrays, the number of antenna units per subarray, and the spacing between antenna units. As the array size increases, the beam becomes more directional and narrower, effectively reducing internal interference and enhancing spectral efficiency \cite{refM004-2}. In the FR-1 band, base stations are typically equipped with 32 to 128 antenna units, whereas in the FR-2 band, higher-density antenna arrays are often required to efficiently support multi-user communications, compensating for the larger path loss. The directionality of an antenna is determined by its subarray structure and directly affects the precise coverage through beamwidth \cite{refM004}.

\subsubsection{Hardware implementation and channel optimization}
In terms of hardware implementation, massive MIMO systems emphasize low-power designs to accommodate multi-user requirements. In the FR-1 band, each antenna unit is typically equipped with low-power, low-noise amplifiers, coupled with digital beamforming technology to optimize user experience. Conversely, in FR-2, HBF is more frequently employed to decrease hardware costs and energy consumption by reducing the number of RF links. The base station adopts the 3GPP CSI acquisition framework, significantly improving beam transmission efficiency through real-time CSI reporting and specific beam-optimization mechanisms, such as TCI or SRS enhancement \cite{refM004-15}. The integration of multi-beam operation and advanced beam management techniques, such as joint transmission and SRS frequency hopping, in millimeter-wave scenarios has been shown to enhance spectrum utilization and expand user coverage \cite{refM004}.

\subsection{Massive MIMO-assisted PLA Classification via Authentication Mechanisms}
This subsection categorizes PLA techniques in massive MIMO systems according to the following criteria: the utilization of channel characteristics for massive MIMO-assisted PLA; tag-based cooperative authentication mechanisms for massive MIMO-assisted PLA; precoder design for massive MIMO-assisted PLA; and addressing hardware impairments in massive MIMO-assisted PLA.

\subsubsection{Utilizing channel characteristics for Massive MIMO-assisted PLA}
In recent years, researchers have proposed various innovative approaches to effectively leverage the complex and rich physical channel characteristics of massive MIMO systems to enhance authentication performance. The following systematically reviews and summarizes representative massive MIMO-assisted PLA schemes that utilize channel characteristics, providing a reference for future research and practical system design.

\begin{itemize}
	\item \textbf{Utilizing channel gain and phase noise:}
	Zhang et al. introduced a novel PLA framework for massive MIMO systems that jointly exploits location-specific channel gains and transmitter-specific phase noise \cite{refM8}. This method employs statistical signal processing tools and matrix analysis to estimate channel gains using linear minimum-mean-square-error (LMMSE) techniques. The LMMSE-based channel estimation produces an estimate $\hat{\boldsymbol{h}}_{\mathrm{AB}}(t)$ that can be expressed as the true channel vector $\boldsymbol{h}_{\mathrm{AB}}(t)$ minus an unknown zero-mean complex Gaussian error vector $\boldsymbol{e}_{h_{\mathrm{AB}}}(t)$, whose covariance matrix captures the combined effects of signal power scaling and additive noise \cite{refM8-45,refM8-36,refM8-46}. The framework also tracks phase noise using an Extended Kalman Filter (EKF), effectively capturing real-time channel variations. By constructing a composite hypothesis-testing model, the researchers derive theoretical expressions for the false-positive and detection probabilities and analyze performance under different channel covariance matrix models. Extensive numerical simulations validate the effectiveness of the proposed framework. Compared to traditional methods that rely solely on channel gains for authentication \cite{refM8-19,refM8-20} or factor in hardware impairments such as phase noise \cite{refM8-22,refM8-23,refM8-25}, the joint utilization of channel gain and phase noise features not only significantly improves system authentication performance but also effectively mitigates the risk of spoofing and replay attacks.
	
	\item \textbf{Deep Learning-assisted Millimeter-wave channel authentication:}
	Zeng et al. introduced a Complex-valued Classification Autoencoder (CVCA) to enhance the effectiveness of PLA against cloning attacks in millimeter-wave (mmWave) massive MIMO communication systems \cite{refM6}. The mmWave channel exhibits significant sparsity due to Line-of-sight (LoS) paths and a limited number of reflected paths. Meanwhile, the high-dimensional nature of massive MIMO systems complicates channel analysis. To address this, the method in \cite{refM6-20} adopts sparse channel modeling based on a virtual channel representation, decomposing the channel matrix via a 2D discrete Fourier transform into a sparse basis set of angular and delay bins. This transformation reveals the inherent sparsity of the mmWave channel and enables efficient feature extraction with reduced computational complexity. The channel fingerprints extracted through this modeling capture the reflection and scattering characteristics of signals at different angles and delays, significantly improving the detection and classification of legitimate nodes. By incorporating complex-valued Long Short-Term Memory modules \cite{refM6-19}, the CVCA processes both the real and imaginary parts of the signal simultaneously, effectively capturing the phase and amplitude characteristics of the channel. This design not only enhances node classification and clone node detection performance but also improves node identification accuracy in multi-antenna systems, as detection probability is positively correlated with the number of antennas.
	
	\item \textbf{Channel geolocation considering hardware impairments:}
	In massive MIMO systems, the impact of hardware impairments on communication performance garners considerable attention \cite{refM8-36,refM5-7}. However, their influence on the PLA is not systematically investigated. Zhang et al. systematically study the impact of hardware impairments on PLA performance in massive MIMO systems and propose a novel authentication scheme \cite{refM5}. This scheme leverages the geographical characteristics of the channel, such as spatial uniqueness and time variability, for authentication while accounting for hardware impairments. Specifically, hardware impairments at the transmitter and receiver are modeled as independent additive distortion noises. The noise power due to transmitter hardware impairments is proportional to the signal power, which can be expressed as
	\begin{equation}
		\boldsymbol{\eta}_t[k] \sim \mathcal{CN}\left(0, \kappa_t p \right),
		\label{eq:tx_hw}
	\end{equation}
	where $\kappa_t$ represents the level of transmitter hardware impairment and $p$ is the average signal power. Similarly, the noise power of the receiver hardware impairment is proportional to the signal power, i.e
	\begin{equation}
		\boldsymbol{\eta}_B[k] \sim \mathcal{CN}\left(0, \kappa_B p \cdot \operatorname{diag}\left(|h_{t1}[k]|^2, \ldots, |h_{tM}[k]|^2\right)\right),
		\label{eq:rx_hw}
	\end{equation}
	where $\kappa_B$ represents the level of receiver hardware impairment and $M$ is the number of receiving antennas. Through theoretical analysis and simulation validation, they derive the PFA and PD under different channel models and explore the impact of hardware impairment levels, signal-to-interference-plus-noise ratio, channel gain ratio ($\gamma$), temporal correlation coefficient ($\alpha$), and the number of antennas ($M$) on authentication performance. The results indicate that hardware impairments significantly degrade authentication performance, whereas increasing the number of antennas can effectively enhance system performance. Moreover, even when the base station is unaware of certain system parameters, the authentication scheme retains some scalability, particularly when the attacker is far from the base station.
	
\end{itemize}

\textbf{\textit{Lessons Learned:}} Channel characteristics in massive MIMO systems each provide distinct authentication signatures. For PLA, it is beneficial to extract and exploit multiple feature dimensions simultaneously. This approach improves robustness over single-feature methods. Computational complexity and practical channel estimation accuracy remain key factors that need to be balanced in system design.

\begin{figure}[t]
	\centering
	\includegraphics[width=0.425\textwidth]{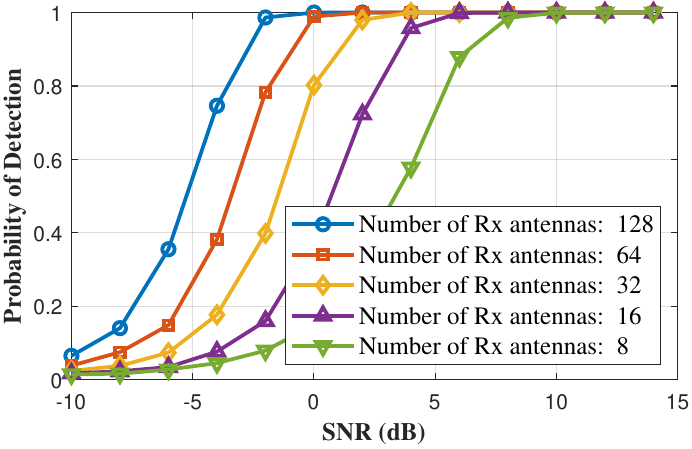}
	\caption{PD versus SNR for massive MIMO-assisted PLA. The results validate channel-based authentication with LMMSE channel estimation using the Frobenius norm distance metric.}
	\label{fig:MassiveMIMOAuth_PD_vs_SNR}
\end{figure}

Fig.~\ref{fig:MassiveMIMOAuth_PD_vs_SNR} provides insights into the performance of massive MIMO-assisted PLA. In this configuration, channel gain characteristics are utilized for user authentication, with the authentication distance metric defined as
\begin{equation}
	D = \sqrt{\sum_{m=1}^{M} |h_{\text{est},m} - h_{\text{ref},m}|^2},
	\label{eq:mimo_distance}
\end{equation}
where $M$ denotes the number of receive antennas.

The simulation uses 10000 Monte Carlo trials. SNR ranges from $-10$ to $15$ dB. The decision threshold is calibrated by a quantile-based procedure to enforce a target PFA of $0.01$. The correlation coefficient between Alice's and Eve's channels is set to $0.6$.
Simulation results reveal that the detection probability increases substantially with the number of antennas. At SNR $= -10$ dB, the PD achieved with $M=8$ antennas is $0.0651$. In contrast, with $M=128$ antennas, the value is $0.9865$, representing a 15.2-fold improvement in performance. At SNR $= 0$ dB, the PD values are $0.2300$ for $M=8$ and $0.9971$ for $M=128$, demonstrating a 4.3-fold performance gain. Across all antenna configurations, near-perfect detection is achieved when SNR $\geq 5$ dB.
These results demonstrate that massive MIMO, through large-scale antenna arrays and spatial diversity, effectively enhances the reliability of PLA.

\begin{figure}[t]
	\centering
	\includegraphics[width=0.425\textwidth]{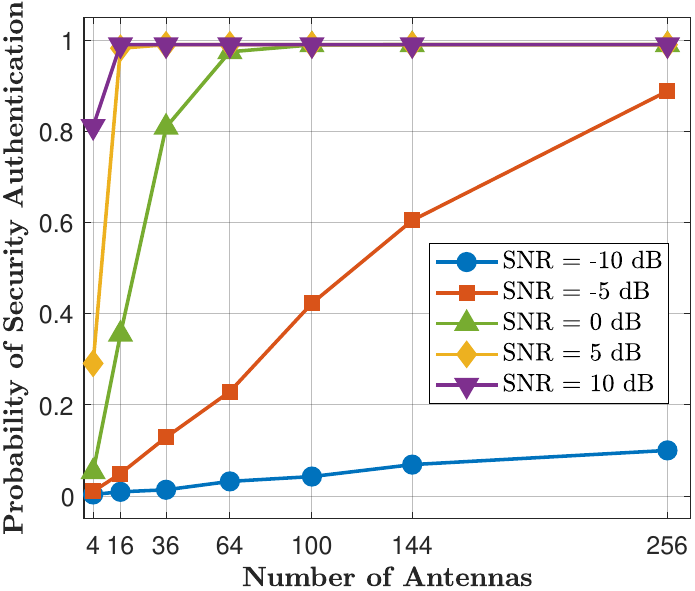}
	\caption{PSA of massive MIMO-assisted PLA versus the number of antennas at Bob. Authentication is based on channel gain features using LMMSE channel estimation.}
	\label{fig:mimo_psa}
\end{figure}

Fig.~\ref{fig:mimo_psa} further evaluates the PSA of massive MIMO-assisted PLA as a function of the number of antennas at Bob. Across all SNR levels, increasing the antenna array size improves the PSA. For example, at SNR $= 0$ dB, the PSA increases from $0.0533$ with 4 antennas to $0.9900$ with 256 antennas. When SNR increases to 5 dB and 10 dB, the PSA also reaches approximately $0.99$ with 256 antennas.

\subsubsection{Tag-assisted cooperative and non-cooperative authentication mechanisms in Massive MIMO}
In massive MIMO systems, tag-assisted PLA mechanisms are categorized into non-cooperative and cooperative user schemes. Non-cooperative schemes primarily rely on tag embedding and the power optimization of single-user signals to achieve efficient authentication, while cooperative schemes further enhance authentication performance through joint multi-user transmission and tag design. The following sections introduce representative methods from both categories and their key technical aspects.

\begin{itemize}
	\item \textbf{No user collaboration:}
	Gu et al. propose a PLA method for incoherent massive SIMO \cite{refM4}, an incoherent massive MIMO technique that effectively improves system reliability and reduces latency in Industrial Internet of Things communication systems through multiple-antenna reception and energy modulation \cite{refM4-5,refM4-6}. Their main contribution is the design of a message-based tag modulation scheme that minimizes the tag Symbol Error Rate (SER) while satisfying the message SER by optimizing the tag embedding scheme and the power allocation between the message and tag signals. Through theoretical analysis and simulation experiments, the results indicate that this method achieves higher authentication accuracy and significantly lower tag SER than the traditional uniform tag embedding scheme \cite{refM4-15,refM4-16,refM4-18} at the same level of communication reliability.
	In \cite{refM3}, the Rayleigh fading channel model is considered, and a linear precoder via zero-forcing precoding \cite{refM3-9} is proposed for data and artificial noise transmission. The authors utilize imperfect location information of an Active Antenna Vehicle Eavesdropper (AAV-Eve) to optimize the precoder design. Through theoretical analysis, they derive closed-form expressions for the ergodic secrecy rate of both conventional and proposed precoder designs. Asymptotic analysis is then employed to identify the optimal power allocation factor to maximize the secrecy rate \cite{refM3-9,refM3-25}.
	
	\hspace*{1em}Furthermore, the proposed precoder design is applied to a fingerprint-embedding authentication framework, optimizing the tag power factor to minimize the probability that an eavesdropper successfully guesses the secret key. Simulation results validate the superiority of the proposed precoder design in terms of secrecy rate and authentication probability, particularly in massive MIMO scenarios with a moderate to large number of antennas. The results demonstrate that even in the presence of angle calibration errors, the proposed precoder design outperforms traditional methods \cite{refM3-10}, effectively reducing the likelihood of successful eavesdropping attacks.
	
	\item \textbf{User collaboration:}
	Unlike the standalone scheme, Yan et al. propose a two-user collaborative PLA scheme for incoherent massive MIMO systems \cite{refM1}. The key to this scheme is the application of a unique constellation decomposition technique \cite{refM1-6} known as the Additively Unique Decomposable Constellation pair (AUDCP), which facilitates efficient communication through space-time modulation \cite{refM1-8}. In this system, the signals from two single-antenna users are transmitted via a specially designed constellation organization that enables noncoherent signal detection in a rapidly changing uplink channel. Specifically, each user transmits a combined signal of its message and an embedded tag simultaneously across two time slots, effectively utilizing both spatial and temporal antenna resources while ensuring high reliability and low latency for message transmission in high-SNR and Rayleigh-fading environments. Additionally, by embedding tags within messages, this scheme enhances protection against replay attacks and spoofed users, thereby further improving the overall security of the system. Simulation results indicate that the authentication performance of this scheme outperforms that of the traditional standalone scheme \cite{refM4}, demonstrating excellent anti-jamming capability and authentication effectiveness.
	
	\hspace*{1em}The study in \cite{refM2} further investigates the tag embedding strategy and its specific impact on authentication performance, and proposes a selective tag embedding mechanism based on dynamic analysis of the positional relationships between messages and labels \cite{refM2-10} to determine the optimal tag embedding scheme. This approach not only enhances label obfuscation but also effectively reduces the risk of malicious users detecting it. Additionally, the article introduces a cooperative working mechanism that combines a minimal Riemann distance detector and a fast noncoherent Maximum Likelihood (ML) detector \cite{refM2-11}. This joint strategy significantly improves label detection and message decoding, enabling the system to maintain effective performance in the presence of noise and interference. The proposed two-user cooperative PLA scheme demonstrates a clear advantage over the reference \cite{refM1} in terms of Bit Error Rate (BER) and detection probability. Through selective tag embedding and noncoherent detection techniques, the scheme presented in this article significantly reduces BER while achieving higher detection probabilities at low false-alarm rates, demonstrating improved authentication performance and robustness.
\end{itemize}

\textbf{\textit{Lessons Learned:}} Tag-assisted authentication in massive MIMO benefits from both user cooperation and noncoherent detection techniques. Optimizing power allocation between message and tag signals, coupled with selective tag embedding, significantly enhances authentication performance, particularly in time-varying channels with fast fading.

\begin{table*}[t]
	\caption{Classification of System Types and Optimization Methods in Technical Papers\label{tab:MIMO_TAG_paper_analysis}}
	\centering
	\renewcommand{\arraystretch}{1.5}
	\begin{tabular}{|m{0.06\textwidth}|m{0.025\textwidth}|m{0.2\textwidth}|m{0.15\textwidth}|m{0.22\textwidth}|m{0.16\textwidth}|}
		\hline
		\textbf{Reference} & \textbf{Year} & \textbf{System Type} & \textbf{Collaboration Method} & \textbf{Tag Embedding Method} & \textbf{Optimization Method} \\
		\hline
		\cite{refM4} & 2018 & Noncoherent massive SIMO industrial IoT system & Joint optimization based on channel characteristics (weak collaboration mechanism) & Non-uniform constellation embedding (dynamically adjusting tag positions based on message symbols) & $\bullet$ Optimal power allocation ratio calculation \newline $\bullet$ Pareto frontier analysis of error performance \\
		\hline
		\cite{refM3} & 2021 & Massive MIMO AAV communication system & No collaboration mechanism (independent authentication based on precoder) & Precoder fingerprint embedding (feature marking through channel precoding) & $\bullet$ Precoding matrix optimization \newline $\bullet$ Artificial noise-assisted fingerprint enhancement \\
		\hline
		\cite{refM1} & 2022 & Noncoherent massive MIMO system & Two-user joint space-time modulation (signal-level collaboration) & Unique decomposed constellation pair (AUDCP) space-time joint modulation & $\bullet$ Approximate symbol error rate analysis \newline $\bullet$ Joint detection threshold design \\
		\hline
		\cite{refM2} & 2023 & Massive MIMO machine-type communications system & Two-user collaboration & Selective tag embedding with bit-level cooperation (based on message-tag bit relationship) & $\bullet$ Minimum Riemann distance detector \newline $\bullet$ Fast noncoherent Maximum Likelihood detection algorithm \\
		\hline
	\end{tabular}
\end{table*}

\subsubsection{Two-stage user authentication framework in Massive MIMO-assisted PLA}
By integrating cryptography with physical layer characteristics, staged authentication is emerging as an effective security mechanism in massive MIMO systems. A typical framework enhances security and resistance to attacks through initial authentication and channel verification.

In the realm of Cellular-V2X (C-V2X) communication security within massive MIMO systems, Kavaiya et al. introduced a two-stage user authentication framework \cite{refM9}. The study specifically analyzed channel characteristics under the Double Generalized Gamma (DGG) fading model for C-V2X scenarios. The DGG model is a flexible statistical distribution characterized by two shape parameters ($m_1$ and $m_2$) and a scale parameter ($a$), which can accurately capture a wide range of fading conditions encountered in vehicular communication environments \cite{refM9-13}.

Grounded in the DGG model, the authentication framework consists of two stages. In the initial authentication stage, users complete registration by providing credentials (such as usernames, passwords, and public keys), and the BS verifies user identities using a challenge-response mechanism. If the initial authentication is successful, the system proceeds to the channel estimation and authentication stage. At this point, the base station sends pilot signals to the user, who estimates the CSI using the received signals. Subsequently, the base station and user conduct secure channel training using the estimated CSI, optimize signal quality through beamforming and precoding, and complete mutual authentication using cryptographic protocols (such as challenge-response or digital signatures). Finally, shared session keys are generated for secure communication via key exchange protocols (such as Diffie-Hellman or RSA). Simulation results indicate that this authentication scheme improves the secrecy rate, particularly as the number of base-station antennas increases, thereby enhancing the spatial diversity of the system and its ability to counter eavesdropping.

\begin{figure*}[t]
	\centering
	\includegraphics[width=0.8\textwidth]{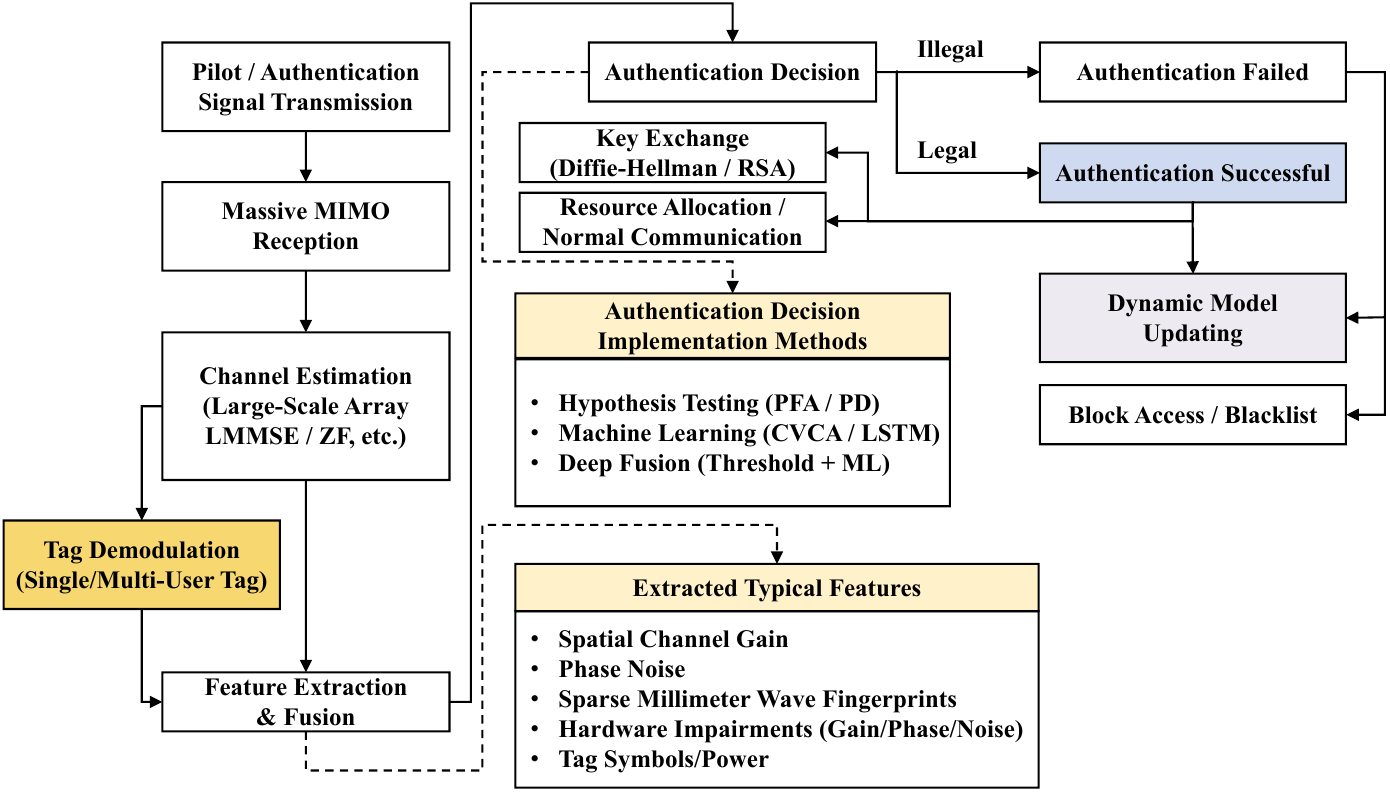}
	\caption{A Universal Integrated Framework for Massive MIMO-assisted PLA.}
	\label{MASSIVE-MIMO-PLA}
\end{figure*}

After systematically surveying various technical routes for massive MIMO-assisted PLA in the preceding subsections, we observe that these approaches share a highly consistent core authentication procedure. Specifically, whether schemes rely solely on channel characteristics, employ complex tag mechanisms, or integrate cryptography into two-stage frameworks, they all fundamentally follow key steps such as signal acquisition, feature extraction, decision strategy, and feedback on authentication results. The differences mainly appear in the details of feature extraction and the emphasis on various decision methods. To intuitively present the intrinsic connections and technical threads among these schemes, this article provides a comprehensive summary of representative studies and further proposes and illustrates a universal, integrated workflow for massive MIMO-assisted PLA applicable across different subclasses, as shown in Fig.~\ref{MASSIVE-MIMO-PLA}.

\begin{table*}[t]
	\caption{Classification of Pilot Contamination Mitigation Techniques\label{tab:mitigation_techniques}}
	\centering
	\renewcommand{\arraystretch}{1.5}
	\begin{tabular}{|m{3.65cm}|m{3.85cm}|m{5.8cm}|m{2.3cm}|}
		\hline
		\textbf{Category} & \textbf{Technique} & \textbf{Mechanism} & \textbf{References} \\
		\hline
		\multirow{2}{*}{Pilot Allocation Optimization} & Dynamic pilot scheduling & Time-shifted pilots, asynchronous transmission & \cite{M008-38, M009-26, M009-27, M009-20, M009-32},\newline \cite{M011-42} \\
		\cline{2-4}
		& Graph coloring algorithm & Minimize pilot reuse distance via graph theory & \cite{M009-31, M011-42} \\
		\hline
		Power Control Technology & Adaptive power control & Dynamic adjustment based on fading coefficients & \cite{M009-30, M010-61} \\
		\hline
		\multirow{2}{*}{Blind Deconvolution Methods} & Covariance-based separation & Statistical channel covariance analysis & \cite{M008-95, M008-103, M008-105, M011-44} \\
		\cline{2-4}
		& Precoding-based decontamination & Spatial suppression using reciprocity & \cite{M008-27, M008-31, M008-98} \\
		\hline
		Hybrid Pilot Design & Superimposed pilots & Pilot-data overlay for resource efficiency & \cite{M011-33} \\
		\hline
		\multirow{2}{*}{Antenna/Hardware Optimization} & Mutual coupling suppression & Nanoantenna arrays with decoupling structures & \cite{M011-10, M011-9} \\
		\cline{2-4}
		& Transceiver calibration & Phase noise and reciprocity compensation & \cite{M010-119, M008-90} \\
		\hline
	\end{tabular}
\end{table*}

\subsection{Potential Security Threats of Massive MIMO-assisted PLA}

Pilot contamination compromises channel estimation accuracy in massive MIMO systems, substantially undermining the robustness of PLA \cite{M008}.

\subsubsection{Security Threats Caused by Pilot Contamination}
Pilot contamination interferes with channel estimation and introduces multi-layered security risks. The following outlines its principles, impacts, and threat scenarios.

\begin{itemize}
	\item Principle of guided contamination attack:
	In time-division duplex massive MIMO systems, attackers transmit identical or related pilot sequences during legitimate users' uplink pilot transmissions \cite{M008-27, M008-32}, causing confusion in channel estimation \cite{M011-34}.
	
	\item Impact on Massive MIMO-assisted PLA:
	\begin{enumerate}[label=\textcircled{\small\arabic*}]
		\item Channel uniqueness compromise: Pilot contamination introduces malicious components, weakening channel discriminability \cite{M008-27, M011-46} and increasing the likelihood of successful impersonation.
		\item Energy efficiency and security trade-off: Additional interference noise forces higher pilot power or authentication thresholds \cite{M009-23}, decreasing energy efficiency and weakening security \cite{M009-30}.
	\end{enumerate}
	
	\item Potential Threat Scenarios:
	\begin{enumerate}[label=\textcircled{\small\arabic*}]
		\item Active jamming attacks: Attackers exploit pilot contamination to interfere with channel feature extraction, breaching channel-based defenses \cite{M008-38}.
		\item Hardware-level vulnerabilities: Hardware defects and transceiver non-ideality \cite{refM8-36} amplify security vulnerabilities caused by pilot contamination.
	\end{enumerate}
\end{itemize}

\subsubsection{Mitigation Methods for Pilot Contamination in MIMO-assisted PLA}
Researchers have proposed various mitigation strategies to address pilot contamination in MIMO-assisted PLA systems:
\begin{itemize}
	\item Pilot allocation optimization:
	\begin{enumerate}[label=\textcircled{\small\arabic*}]
		\item Dynamic pilot scheduling: Time-shifted pilots \cite{M008-38, M009-26, M009-27} or asynchronous transmission \cite{M009-20, M009-32} reduce multi-cell pilot collisions \cite{M011-40}.
		\item Graph coloring algorithm: Pilot allocation is modeled as a graph coloring problem to minimize pilot reuse distance \cite{M009-31, M011-42}.
	\end{enumerate}
	
	\item Power control:
	Adaptive pilot power control adjusts power according to large-scale fading coefficients to suppress long-range interference \cite{M009-30, M010-61}.
	
	\item Blind deconvolution (subspace techniques):
	Covariance-based blind separation uses channel covariance statistics to separate legitimate users from contaminating signals \cite{M008-95, M008-103, M008-105, M011-44}. Precoding-based decontamination designs precoding matrices based on channel reciprocity to suppress contaminating signal propagation \cite{M008-27, M008-31, M008-98}.
	\item Hybrid pilot design:
	Superimposed pilots overlay pilot signals onto data symbols, reducing dedicated pilot resource overhead \cite{M011-33}.
	\item Antenna and hardware optimization:
	Mutual-coupling-suppression antenna designs \cite{M011-10, M011-9} and transceiver calibration techniques \cite{M010-119, M008-90} mitigate contamination from hardware impairments.
\end{itemize}

These mitigation techniques are classified and compared in Table \ref{tab:mitigation_techniques}.

\subsection{Summary} 
We present massive MIMO hardware architectures, fundamental principles of spatial multiplexing and beamforming, authentication mechanisms via channel characteristics, tag-assisted schemes, and pilot contamination challenges. The advantages of massive MIMO-assisted PLA stem from the high spatial resolution and degrees of freedom provided by large-scale antenna arrays, which enables strong extraction of high-dimensional spatial features. However, pilot contamination emerges as a critical bottleneck that compromises channel estimation robustness and authentication reliability. Future designs should jointly address robust pilot design, contamination detection, and lightweight authentication strategies to balance security with system efficiency.


\section{Space-domain Programmable RIS for PLA}
\label{sec5}
This section examines how RIS enhances PLA by enabling space-domain programmability. Building upon the previous evolution of space-domain enhancement technologies, such as DPA for internal polarization-domain distinguishability and massive MIMO for internal spatial sampling and high spatial resolution, RIS enables external propagation-environment reconfiguration and further extends the notion of spatial enhancement from exploiting richer internal channel structures to actively shaping the external propagation mapping.  We begin by introducing the hardware architecture, operational principles, and channel models of RIS in RIS-assisted PLA. Subsequently, we categorize and examine four major RIS-assisted authentication mechanisms. Finally, we discuss the security challenges inherent to RIS-assisted PLA and summarize defensive measures.
\begin{table*}[t]
	\caption{Summary of RIS-assisted PLA Channel Types.\label{RIS_channel_type}}
	\centering
	\renewcommand{\arraystretch}{1.5}
	\begin{tabular}{|c|m{12cm}|m{1.5cm}|}
		\hline
		\textbf{Channel Type} & \textbf{Literature Summary} & \textbf{References} \\
		\hline
		\multirow{11}{*}{\textbf{Cascaded channel}} & Using MATLAB and SimRIS to construct a RIS wireless communication model. By optimizing the phase, the PLA performance via path loss and CIR was significantly improved. & \cite{ref69} \\
		\cline{2-3}
		& BS-RIS-AAV: BS-RIS is a quasi-static channel, whereas RIS-VAV is a time-varying channel due to AAV mobility. & \cite{ref66} \\
		\cline{2-3}
		& GCS-RIS-AAV: The cascaded channels are modeled as equivalent point-to-point Nakagami fading channels. & \cite{ref71} \\
		\cline{2-3}
		& All channels are time-invariant, but the configuration of the RIS is controlled by Bob and can be modified over time, thereby rendering the cascaded channel both time-varying and controllable. & \cite{ref76}, \cite{ref73} \\
		\cline{2-3}
		& Enhance the power of legitimate signals and optimize transmission through RIS, while using CFO for authentication, combining it with CSI to optimize RIS configuration in order to achieve the best authentication results under confidentiality constraints. & \cite{refR20} \\
		\hline
		\multirow{9}{*}{\textbf{Combined channel}} & RIS enhances CIR and then uses a Convolutional Neural Network (CNN) for identity recognition, periodically updating the legitimate CIR database. & \cite{ref72} \\
		\cline{2-3}
		& The introduction of the RIS reduces the overlap of the RSS probability density functions between Alice and the eavesdroppers. & \cite{ref74} \\
		\cline{2-3}
		& The introduction of additional multipath and dual-structure sparsity in cascaded channels is proposed for RIS-MIMO systems. & \cite{ref70} \\
		\cline{2-3}
		& The channels are estimated in each coherent block using the LS algorithm during the authentication phase, and the estimation error covariance matrix is determined for performance evaluation. & \cite{ref67} \\
		\cline{2-3}
		& Quasi-static block fading model. Constructing tag signals by leveraging RIS-related wireless channel characteristics, background noise, and random signals. & \cite{ref65} \\
		\hline
	\end{tabular}
\end{table*}

\begin{figure}[t]
	\centering
	\includegraphics[width=3.4in]{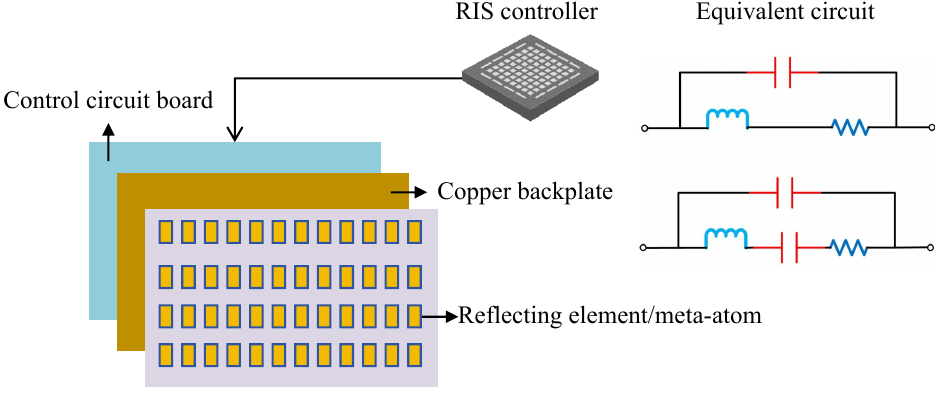}
	\caption{Hardware Architecture of RIS.}
	\label{Hardware Architecture of RIS}
\end{figure}

\begin{figure*}[t]  
	\centering  
	\includegraphics[width=0.85\textwidth]{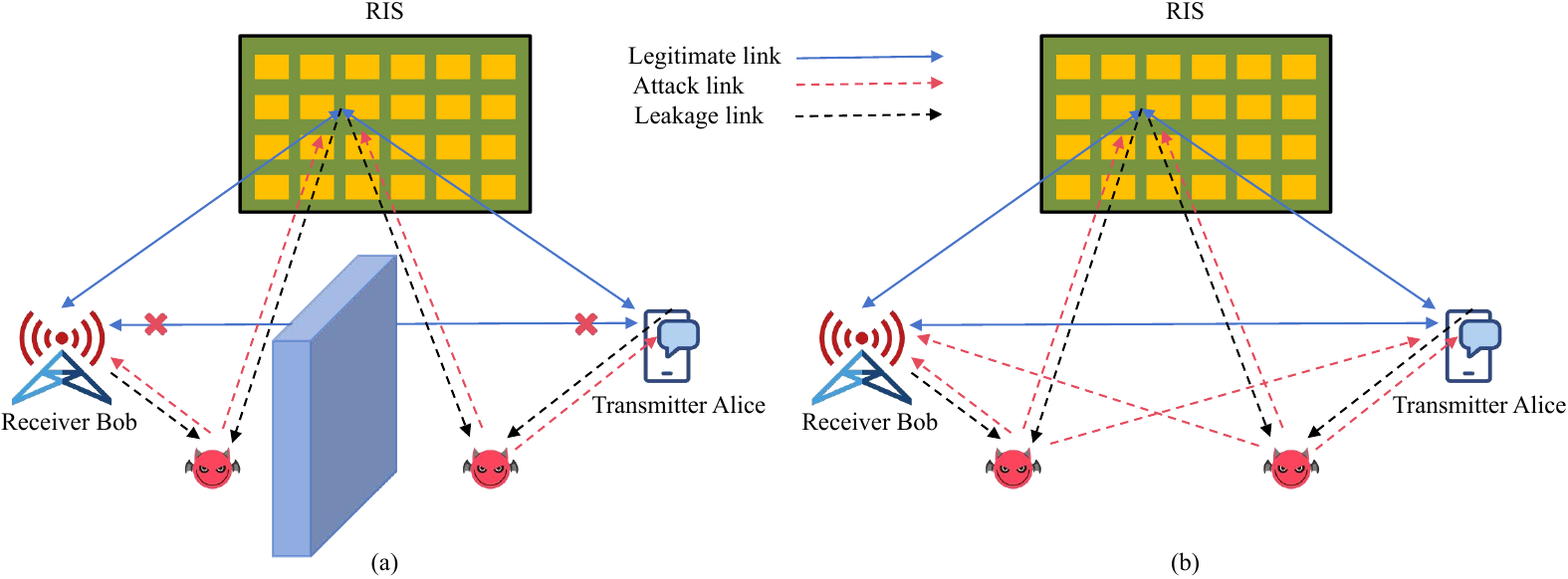} 
	\caption{RIS-assisted PLA system models under different channel types: (a) cascaded channel and (b) composite channel, each with eavesdroppers positioned near both Alice and Bob.}
	\label{PLA system models under different channel types}  
\end{figure*}  

\subsection{Basic Principles and Characteristics of RIS Technology}
A key advantage of RIS is its precise phase and amplitude control, which enables fine-grained manipulation of reflected signals. The typical hardware architecture of RIS consists of three layers and an intelligent controller, as illustrated in Fig.~\ref{Hardware Architecture of RIS}. The outer layer comprises numerous metal patches printed on a dielectric substrate that interact directly with the incoming signal. Behind this layer, copper or other metal plates prevent signal energy leakage. The inner layer consists of the control circuit board, which regulates the reflection amplitude and phase offset of each element and is driven by the intelligent controller connected to the RIS. These components together realize precise three-dimensional reflection beam formation, whose configuration can be dynamically adjusted \cite{risbeam1}.

Unlike adaptive techniques in conventional transceiver links, RIS can dynamically reconfigure the reflected signal path under the control of the receiver to reconstruct the wireless propagation environment. This controllable reflection characteristic provides a new dimension for PLA, since the control authority over the reflection configuration at the receiver enables it to be leveraged as an authentication feature \cite{risbeam2}. The high-dimensional feature space created by phase and amplitude control across multiple RIS elements provides a rich foundation for implementing fine-grained physical layer authentication mechanisms \cite{risbeam3}.

\subsection{Cascaded Channel and Composite Channel}
RIS-assisted PLA considers various channel types, including cascaded channels and their combinations with direct channels. We refer to this combination of cascaded and direct channels as a composite channel.

\subsubsection{Cascaded channel}
In a typical RIS configuration, the direct channel between the legitimate communicating parties is often blocked \cite{ref68}. Fig. \ref{PLA system models under different channel types} (a) illustrates the cascaded channel model of a PLA system, where the cascaded channel denotes the signaling paths from Alice to RIS and from RIS to Bob. The cascaded channel can be represented as
\begin{equation}
	h_{\text{cascaded}}=\sum_{n=1}^N h_{\mathrm{AR}, n}  \theta_n  h_{\mathrm{RB}, n},
	\label{eq:cascaded_channel}
\end{equation}
where $N$ represents the total number of reflection elements in the RIS; $h_{\mathrm{AR}, n}$ is the channel coefficient from Alice to the $n$-th element of the RIS; $\theta_n=e^{j \phi_n}$ denotes the reflection coefficient of the $n$-th element of the RIS; $\phi_n$ is the adjustable phase shift; $h_{\mathrm{RB}, n}$ represents the channel coefficient from the $n$-th element of the RIS to the receiver. By adjusting the phase shift $\phi_n$ of each RIS element, the phase of the reflected signal can be controlled, thereby enhancing the desired signal or suppressing interference, improving the performance of PLA. The cascaded channel $h_{\text{cascaded}}$ captures the role of the RIS in channel regulation, a key distinction between RIS-assisted communication systems and traditional systems. 

\subsubsection{Composite channel}
Fig. \ref{PLA system models under different channel types} (b) depicts the composite channel PLA system model. The composite channel RIS-assisted PLA system comprises a direct path from the transmitter to the receiver and a reflected path via the RIS, which can be given by
\begin{equation}
	\begin{aligned}
		h_{\text{composite}} &= \{h_{\text{direct}}, h_{\text{cascaded}}\}, \\
		&= \left\{h_{\text{direct}}, \sum_{n=1}^N h_{\mathrm{AR},n} \theta_n h_{\mathrm{RB},n}\right\},
	\end{aligned}
	\label{eq:composite}
\end{equation}
where $h_{\text{direct}}$ is the direct channel coefficient between the transmitter and receiver. The composite channel $h_{\text {composite }}$ combines the effects of the direct and reflected paths and is the basis for signal processing and authentication at the receiver.  We summarize the channel types discussed in RIS-assisted PLA literature in Table \ref{RIS_channel_type}.

\textbf{\textit{Lessons Learned:}} Channel modeling is foundational to RIS-assisted PLA design. The choice between cascaded and composite channels directly determines the receiver-side observation model and, in turn, the separability of authentication features. Therefore, proper channel characterization is essential for effectively exploiting RIS programmability in authentication-related signal processing.
\begin{table*}[t]
	\caption{Utilization of Channel Characteristics in RIS-assisted PLA\label{RIS_channel}}
	\centering
	\renewcommand{\arraystretch}{1.3}
	\begin{tabular}{|l|l|l|}
		\hline
		\textbf{Type} & \textbf{Channel Characteristics} & \textbf{References} \\
		\hline
		\multirow{2}{*}{\textbf{Complete channel information}} & Channel state information& \cite{ref68} \\
		\cline{2-3}
		& Channel impulse response & \cite{ref69}, \cite{ref72} \\
		\hline
		\multirow{2}{*}{\textbf{Energy-related characteristics}} & Channel gain and path loss & \cite{ref66}, \cite{ref69} \\
		\cline{2-3}
		& Received signal strength/ Energy measurement & \cite{ref71}, \cite{ref74} \\
		\hline
		\multirow{2}{*}{\textbf{Statistical properties}} & Channel covariance matrix & \cite{ref67} \\
		\cline{2-3}
		& Square of channel gain & \cite{ref67} \\
		\hline
		\multirow{2}{*}{\textbf{Structural characteristics}} & MIMO channel sparsity & \cite{ref70} \\
		\cline{2-3}
		& Cascaded channel sparsity & \cite{ref70} \\
		\hline
	\end{tabular}
\end{table*}

\subsection{RIS-assisted PLA Classification via the Authentication Mechanism}
Following the channel classification for RIS-assisted PLA, we further categorize the schemes by the primary source of authentication evidence provided by RIS. The main categories include: utilizing channel characteristics for RIS-assisted PLA; challenge-response mechanism for RIS-assisted PLA (CR-RISAuth); tag utilization method for RIS-assisted PLA; and RIS-assisted PLA leveraging other innovative architectures, where the key contribution comes from architectural/functional augmentation of RIS rather than only the decision rule or protocol form.

\subsubsection{Utilizing channel characteristics for RIS-assisted PLA}
This approach focuses on leveraging the unique characteristics of communication channels for authentication. The fundamental concept underlying this approach is to exploit the programmable attributes of RIS to enhance the complexity of the channel and information integrity by modifying its electromagnetic environment. As summarized in Table \ref{RIS_channel}, the utilization of channel characteristics for RIS-assisted PLA can be categorized into four distinct dimensions: channel response matching, energy characteristics of the signal, sparsity of channel angles, and statistical properties of the channel.

\begin{itemize}
	\item \textbf{Channel response matching:}
	Amin et al. investigate a PLA mechanism for an RIS system that exploits Channel Impulse Response (CIR) properties via channel response matching \cite{ref69}. Their exhaustive search for the optimal phase offset significantly reduces the PMD, particularly within the phase component. By configuring the RIS with this optimal phase offset, the researchers ensure that the PMD remains zero, regardless of fluctuations in link quality or detection thresholds. A subsequent study in \cite{ref72} explores the potential of RISs to augment CIR complexity for data classification and authentication. The RIS modulates wireless signals to improve CIR robustness. A convolutional neural network at the receiver is employed to classify CIR-derived data, while a dynamic method updates the CIR database for authorized transmitters, mitigating environmental impacts on authentication. Experiments with the DeepMIMO dataset \cite{refR10-18} and ray-tracing simulations demonstrate that RIS significantly improves detection rates for both legitimate and illegitimate sender nodes, especially during initial attacks from unauthorized nodes.
	
	\noindent For CIR-based RIS-assisted PLA, Bob observes the cascaded channel response $h_{\text{cascaded}} = \sum_{n=1}^{N} h_{\mathrm{AR},n}\theta_n h_{\mathrm{RB},n}$ (as defined in~\eqref{eq:cascaded_channel}) shaped by the RIS phase configuration $\boldsymbol{\theta} = \{\theta_1,\dots,\theta_N\}$, where $h_{\mathrm{AR},n}$ and $h_{\mathrm{RB},n}$ denote the Alice-RIS and RIS-Bob channel coefficients, respectively. The RIS phase shifts $\theta_n$ are configured to an optimal setting determined via exhaustive search, or alternatively left as controllable degrees of freedom for CIR augmentation.
	
	\item \textbf{Energy characteristics of the signal:}
	To address the challenges of cascaded channel authentication in RIS-assisted AAV systems, Qin et al. focus on two key characteristics: the time-varying nature of the RIS-AAV channel and the quasi-static property of the BS-RIS channel \cite{ref66}. They choose the Least Squares (LS) algorithm for estimating the RIS-AAV channel, considering its suitability for handling short-term variations \cite{refR02-40}. For the quasi-static BS-RIS channel, they employ the coordinate descent method, which decomposes the problem into subproblems for estimating the channel coefficients of specific RIS elements independently \cite{refR02-34}. Continuous channel gain values are quantized into discrete values \cite{refR02-20}, with the output of a 1-bit quantizer used to generate authentication decision criteria for binary hypothesis testing \cite{refR02-19}. Alternatively, Wang et al. propose a unique authentication method by measuring received signal energy for binary hypothesis testing \cite{ref71}. This involves deriving the PDF of the signal energy, enabling authentication by comparing a detection metric to a threshold to determine if a signal comes from an authorized AAV.
	
	\hspace*{1em}Meanwhile, optimizing path loss in RIS systems is crucial. Amin et al. use an exhaustive search to minimize undetected path loss by finding the optimal phase shift, revealing that while RIS and non-RIS systems show similar PFA, optimizing RIS phase shifts significantly decreases PMD, even in the presence of channel noise \cite{ref69}. Furthering this work, Gao et al. develop a prototype system for RIS-assisted Signal Emitter Identification (SEI) to address the need for spatial separation of transmitters \cite{ref74}. Their system allows a valid transmitter to alter the channel fingerprint during SEI by toggling the RIS. Theoretical analyses suggest that using RIS effectively reduces the overlap of the Received Signal Strength (RSS) probability density functions of Alice and Eve, indicating that RIS adds an intrinsic property reflected in the mean and variance of the RSS distribution.
	
	\noindent For energy-characteristic-based RIS-assisted PLA, the baseband observation at Bob follows the cascaded channel model~\eqref{eq:cascaded_channel} (when the direct link is blocked) or the composite channel model~\eqref{eq:composite} (when both direct and reflected links exist). The received signal power $|y_{\mathrm{B}}|^2$ or energy $E_{\mathrm{B}}$ serves as the observable metric for authentication.
	
	\item \textbf{Sparsity of channel angles:}
	In MIMO systems, it has been observed that the supplementary multipath components introduced by RIS result in a virtual channel representation that is angularly sparse \cite{ref70}. This sparsity is evident in the distribution of angles of arrival and departure within the virtual channel space. This manifests not only in the limited number of signal propagation paths but also in the distinct sparse properties of the rows and columns of the signal in this space. Specifically, the row support (arrival angle) remains consistent across all users, whereas the column support (departure angle) varies for individual users. This unique dual-structure sparsity facilitates the derivation of authentication signatures by incorporating both direct and cascaded link-angle indices from the RIS-assisted channel. Mathematically, this is obtained by projecting the spatial channel into the angular domain via unitary discrete Fourier transform matrices, where the RIS phase configuration matrix $\boldsymbol{\Theta} = \operatorname{diag}(\theta_1,\dots,\theta_N)$ shapes the contribution of the cascaded link.
	
	\item \textbf{Statistical properties of the channel:}
	The dynamic alterations in the channel induced by RIS deployment, including increased channel complexity and reduced temporal correlation, are highlighted in \cite{ref67}. Moreover, the study leverages the second-order channel statistics and the channel covariance matrix (CCM) as auxiliary information. By comparing the squared values of the extracted channel against predefined thresholds and incorporating the stability of the CCM, a multi-detector framework based on the likelihood ratio test is designed to effectively distinguish legitimate users from potential attackers.
	
	\noindent For statistical-property-based RIS-assisted PLA, the signal model follows the composite channel structure~\eqref{eq:composite}. The channel covariance matrix $\boldsymbol{R}_{h} = \mathbb{E}[\boldsymbol{h}_{\text{ch}} \boldsymbol{h}_{\text{ch}}^{\dagger}]$, which captures the second-order statistics of the composite channel, serves as the key feature for the multi-detector framework under the quasi-static block fading assumption.
\end{itemize}

\subsubsection{Challenge-response Mechanism for RIS-assisted PLA (CR-RISAuth)}
This approach employs an interactive method in which challenges are issued, and responses are evaluated to verify authenticity. The core idea is that when the receiver controls the RIS, it can apply random configurations that remain secret from the attacker. This allows the receiver to confirm whether the estimated channel from the received signal aligns with the predicted channel via the applied configuration.

\begin{algorithm}[t]
	\caption{Algorithm for CR-RISAuth}
	\label{alg:CR-RISAuth}
	\begin{algorithmic}[1]
		\STATE \textbf{Input:} Alice (transmitter), Bob (receiver), RIS with $N$ reflection elements.
		\STATE \textbf{Output:} LEGITIMATE or ILLEGAL.
		\STATE \textbf{Notation:} $\delta$ is the authentication threshold; $h_{\text{est}}$ denotes the estimated channel state; $h_{\text{exp}}$ is the expected channel state.
		\STATE \textbf{Phase 1: CSI Measurement and Identification}
		\STATE Alice transmits authenticated pilot signals to Bob over the RIS-assisted channel under $K$ known RIS configurations $\{\boldsymbol{\theta}^{(1)},\dots,\boldsymbol{\theta}^{(K)}\}$.
		\STATE For each configuration $\boldsymbol{\theta}^{(k)}$, Bob estimates and stores the cascaded channel state information $\text{CSI}_{\text{stored}}(\boldsymbol{\theta}^{(k)})$, enabling prediction of the cascaded channel for any RIS configuration in Phase 4.
		\STATE \textbf{Phase 2: Challenge (Random RIS Configuration)}
		\STATE Bob randomly selects a new RIS configuration $\boldsymbol{\theta} = \{\theta_1,\theta_2,\dots,\theta_N\}$ as the implicit challenge.
		\STATE Only Bob knows the current RIS configuration (challenge is confidential).
		\STATE \textbf{Phase 3: Message Transmission}
		\STATE Alice transmits message $m$ to Bob through the RIS (without knowledge of Bob's random configuration).
		\STATE Bob receives the signal $y_{\mathrm{B}} = \bigl(\sum_{n=1}^{N} h_{\mathrm{AR},n}\theta_n h_{\mathrm{RB},n}\bigr) \, m + n_{\mathrm{B}}$ under the current RIS configuration $\boldsymbol{\theta}$.
		\STATE Bob estimates the cascaded channel from the received signal: $h_{\text{est}} = \sum_{n=1}^{N} h_{\mathrm{AR},n}\theta_n h_{\mathrm{RB},n}$.
		\STATE \textbf{Phase 4: Channel Verification}
		\STATE Using $\text{CSI}_{\text{stored}}$ (from Phase 1), Bob computes the expected cascaded channel for the chosen configuration $\boldsymbol{\theta}$, i.e., $h_{\text{exp}} = \text{CSI}_{\text{stored}}(\boldsymbol{\theta})$.
		\IF{$|h_{\text{est}} - h_{\text{exp}}| < \delta$}
		\STATE Output: \textbf{LEGITIMATE}.
		\ELSE
		\STATE Output: \textbf{ILLEGAL}.
		\ENDIF
	\end{algorithmic}
\end{algorithm}

Tomasin et al. first explored the potential of partially controllable channels in PLA, proposing novel authentication mechanisms called CR-RISAuth \cite{ref24}. This concept achieves authentication by altering the physical characteristics of the electromagnetic environment. The fundamental idea is that the receiving device controls the channel configuration to pose an implicit challenge, and the response is the cascaded channel estimated from the received signal. By checking whether the estimated channel matches the expected channel under the chosen configuration, Bob can effectively prevent malicious attackers from forging responses. Simulation results show that increasing the number of RIS configurations significantly reduces PMD and Eve's success rate \cite{ref82}. Even with the attacker's ability to obtain perfect Channel State Information (CSI) for both the Alice-to-RIS and RIS-to-Bob channels, the security performance remains superior to that of traditional tag authentication. Subsequently, Tomasin et al. apply the concept of CR-RISAuth to secure communication in cellular systems, demonstrating that a good balance between security and communication performance can be achieved by judiciously selecting the RIS randomness parameter \cite{ref76}. Here, guided by insights from prior research, we present the general flow of the CR-based mechanism in RIS-assisted PLA, comprising four main steps, as detailed in Algorithm \ref{alg:CR-RISAuth}.

Specifically, Guglielmi et al. advance Tomasin et al.'s work by designing a CR-RISAuth system that integrates RIS with a single-antenna device to improve the trade-off between communication performance and security \cite{ref73}. They introduce a probability distribution design via random RIS configurations to optimize system performance while adhering to constraints on PFA and PMD, thereby maximizing the average received SNR. Simulation results indicate that increasing SNR reduces PMD but increases PFA.

\noindent In CR-RISAuth, the received signal at Bob follows the cascaded channel model under a random RIS configuration $\boldsymbol{\theta}$ chosen by Bob (as described in Algorithm~\ref{alg:CR-RISAuth}, Phase 3). The authentication decision is based on comparing the estimated and expected cascaded channels. For our subsequent case study, we define the authentication distance as
\begin{equation}
	D = |h_{\text{est}} - h_{\text{exp}}|,
\end{equation}
where $h_{\text{est}}$ denotes the estimated cascaded channel from the received signal and $h_{\text{exp}}$ denotes the corresponding expected cascaded channel computed from the stored CSI under the applied RIS configuration.

\begin{figure}[t]
	\centering
	\includegraphics[width=0.425\textwidth]{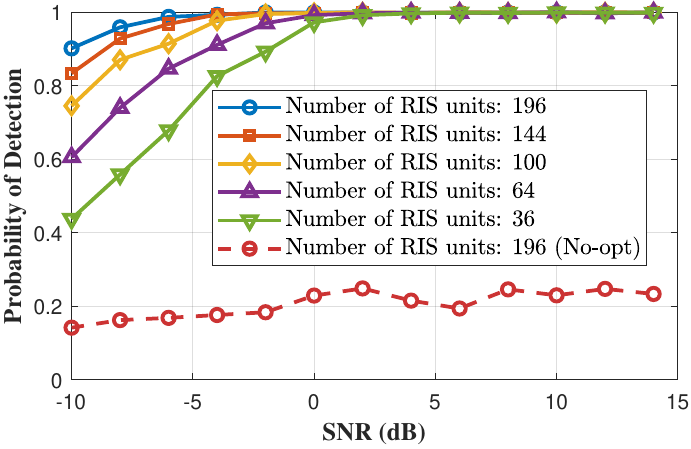}
	\caption{PD versus SNR for CR-RISAuth in RIS-assisted PLA. The results demonstrate the effectiveness of Algorithm~\ref{alg:CR-RISAuth} via choose-best RIS phase optimization over cascaded channels.}
	\label{fig:CR-RISAuth_PD_vs_SNR}
\end{figure}

The RIS configuration is optimized through a discrete-phase random search combined with a minimum-distance criterion, which implements a choose-best phase optimization in the cascaded channel. All simulations adopt unified parameters. 10000 Monte Carlo trials are conducted, and the target false alarm probability is set to $\mathrm{PFA}=0.01$ via quantile-based threshold calibration. The correlation coefficient between Alice's and Eve's channels is set to $0.6$.

Fig.~\ref{fig:CR-RISAuth_PD_vs_SNR} presents the PD versus SNR for different numbers of RIS units. The results show that, under a fixed receiver noise variance, PD improves consistently as the number of RIS units increases, demonstrating that RIS programmability enhances PLA reliability by reshaping the propagation-induced channel realization and thus strengthening the discrepancy captured by the Euclidean-distance-based metric. For instance, at $\mathrm{SNR}=-2$ dB, PD reaches $0.9998$ with 196 RIS units, $0.9955$ with 100 RIS units, and $0.8931$ with 36 RIS units. Moreover, the choose-best phase optimization further improves PD compared with the non-optimized baseline even for large RIS sizes. The optimized configuration achieves a clearly higher PD than the non-optimized baseline and demonstrates that RIS phase design directly improves the discriminability of the authentication distance. Achieving $\mathrm{PD}>0.98$ with 36 RIS units requires only about 2 dB of additional SNR. This result shows that the programmable environment can substantially reduce the sensing and decision burden under moderate channel conditions.

\begin{figure}[t]
	\centering
	\includegraphics[width=0.4\textwidth]{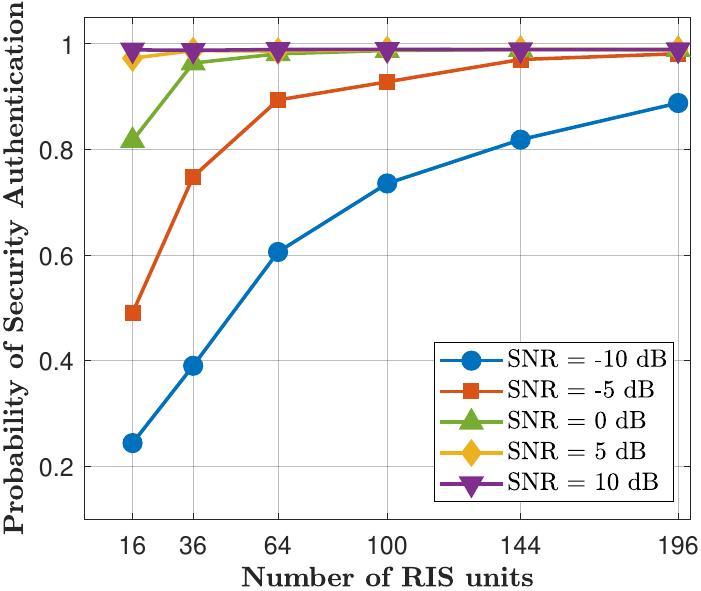}
	\caption{PSA of RIS-assisted PLA (CR-RISAuth) versus the number of RIS reflecting elements. The RIS configuration is optimized via a discrete-phase random search with a choose-best strategy over cascaded channels.}
	\label{fig:ris_psa}
\end{figure}

Fig.~\ref{fig:ris_psa} illustrates the PSA of CR-RISAuth as a function of the number of RIS reflecting elements at different SNR levels. The results show that increasing the number of RIS elements improves the PSA, particularly at low SNR. For example, at SNR = $-10$~dB, the PSA increases from $0.3904$ without RIS to $0.8881$ with 196 reflecting elements, corresponding to a 2.27-fold improvement. At higher SNR levels, such as $0$~dB and above, the PSA is already close to $0.99$ even without RIS, leaving limited room for further enhancement.

\textbf{\textit{Lessons Learned:}} This case study supports the survey perspective that CR-based RIS assistance improves authentication security and reliability by exploiting receiver-controlled random RIS configurations, while appropriate phase optimization further enhances the effectiveness of the distance-based verification metric.

\subsubsection{Tag Utilization Method for RIS-assisted PLA}
This method leverages specific tags for identification, providing a distinctive approach to authentication. Existing tag-based active authentication methods significantly enhance identity verification by detecting tag signals using a message signal and a shared key. However, they are vulnerable to impersonation attacks, as adversaries can forge tags by observing the message signal. Additionally, these methods do not secure tag signals during transmission, exposing them to unauthorized detection, tampering, and man-in-the-middle attacks. The unique characteristics of heterogeneous RIS-assisted systems, such as frequent transmitter switching, also complicate the application of these active schemes to RIS-assisted PLA systems.

Our survey indicates a paucity of studies employing tag-based authentication schemes in RIS-assisted PLA systems. Zhang et al. use the physical-layer attributes of the RIS system, including channel gain and background noise, for authentication \cite{ref65}. They place significant emphasis on the design of tag signals and on the use of asymmetric encryption techniques to ensure their security during transmission. Compared with tag-based PLA schemes that do not employ the RIS, this approach leverages the RIS's physical-layer characteristics for authentication, meticulously designs the tag signals, and encrypts them for transmission using asymmetric encryption. This method offers enhanced security and compatibility but may result in higher computational complexity and performance overhead. In particular, Alice estimates the channel coefficients and background noise for the Alice-RIS and Alice-Bob links using a low-overhead dual time scale channel estimation method, which is denoted as
$h_1^{\left(k_2\right)}, h_d^{\left(k_2\right)}$, and $w^{\left(k_2\right)}$. From the estimation results of Alice-RIS, $\left\{h_{1,1}^{\left(k_2\right)}, \ldots, h_{1, N}^{\left(k_2\right)}\right\}(j \in\{1, \ldots, N\})$, one channel coefficient is randomly selected (both Alice and Bob are equipped with single antennas) and added to the estimation result of Alice-Bob to form a composite channel. Subsequently, the composite channel, combined with a finite discrete symbol set and background noise, generates the labeled signal, which can be represented mathematically as
\begin{equation}
	t_a=\left(h_d^{\left(k_2\right)}+h_{1, j}^{\left(k_2\right)}\right) s+w^{\left(k_2\right)} ,
	\label{eq:tag_signal}
\end{equation}
where the moment $k_2$ represents the subsequent moment of the moment $k_1$, and the construction of the aforementioned tag is carried out at the moment $k_2$. At the moment $k_1$, the identity interaction and initialization exchange are conducted between Alice and Bob, and the private key and public key are distributed to Alice and Bob. To further enhance the security of the tag signal during transmission, asymmetric encryption is applied. The receiver utilizes the maximum a posteriori probability criterion to make authentication decisions. By comparing the received tag signals with the reference tag signals transmitted during the training phase and accounting for the distributions of channel gain, background noise, and random signals, the receiver can determine whether the received signals are legitimate.

\noindent After Alice transmits the encrypted tag signal through the RIS-assisted composite channel, the received signal at Bob follows the composite channel model~\eqref{eq:composite} with the tag signal $t_a$ as the transmitted symbol, where the composite channel combines the direct Alice-Bob path and the cascaded Alice-RIS-Bob path.

In summary, whether it is RIS-assisted PLA or PLA without RIS assistance \cite{refQ0201, refXIE-Hybrid, refXIE-Privacy}, we describe that existing tag-based PLAs typically involve the following steps:
\begin{enumerate}[label=\textcircled{\small\arabic*}]
	\item \text{Initialization:} Alice and Bob exchange identity information or initialization parameters over secure channels to ensure consistent initial information:
	\begin{itemize}
		\item Shared key: Pre-distribute symmetric shared keys for tag generation and authentication.
		\item Channel parameter initialization: Distribution of channel estimates and background noise statistics.
		\item Encryption parameters: Public and private keys may need to be distributed if key pairs are used.
	\end{itemize}
	
	\item \text{Tag generation:} Alice generates a unique tag based on shared keys, channel characteristics, and noise:
	\begin{itemize}
		\item Construct unique labeling signals using random number generating functions.
		\item Apply specific transformation functions according to the channel response.
		\item Combine message signals with secure hash functions to associate labels with transmitted content.
		\item Optionally encrypt the tag using symmetric or private keys to improve security and obtain a cover tag signal.
	\end{itemize}
	
	\item \text{Tag embedding:} The generated tags are embedded into the source signal:
	\begin{itemize}
		\item Embedding can be done through direct superposition or signal modulation.
		\item Balance between signal transmission quality and tag robustness must be maintained.
	\end{itemize}
	
	\item \text{Signal transmission:} The tag-embedded signal is transmitted over the wireless channel to Bob. The signal may suffer from interference due to the random nature of the wireless channel.
	
	\item \text{Tag verification:} Upon reception, Bob verifies the authenticity of the tag:
	\begin{itemize}
		\item Tag extraction: Extract the original tag signal using the received composite signal and the public/shared key, demodulating the overlay tag signal in conjunction with channel characteristics.
		\item Statistical validation: Use statistical methods to assess tag legitimacy, drawing on predefined tag distribution models, channel gain characteristics, and background noise.
		\item Threshold determination: Perform statistical discrimination by setting thresholds upon validation results.
	\end{itemize}
\end{enumerate}

\begin{table*}[t]
	\caption{Comparison of Different Mechanisms of RIS-assisted PLA\label{tbl5}}
	\centering
	\renewcommand{\arraystretch}{1.5}
	\begin{tabular}{|p{0.07\textwidth}|p{0.15\textwidth}|p{0.4\textwidth}|p{0.25\textwidth}|}
		\hline
		\textbf{Mechanism} & \textbf{Classification} & \textbf{Role of RIS} & \textbf{Challenge} \\
		\hline
		\multirow{2}{*}{Channel} & \multirow{2}{*}{Passive, Non-interactive} & Enhance channel characteristic richness and optimize channel features to improve authentication performance. & \multirow{2}{*}{Channel environment changes.} \\
		\hline
		\multirow{2}{*}{CR} & \multirow{2}{*}{Passive, Interactive} & Dynamically adjust RIS configuration to encode challenge information. & Real-time channel estimation and RIS configuration updates. \\
		\hline
		Tag & Active, Interactive & Improve coverage and compatibility of tagged signals. & Tag design and secret key security. \\
		\hline
	\end{tabular}
\end{table*}

\textbf{\textit{Lessons Learned:}} Different RIS-assisted PLA mechanisms present distinct trade-offs between implementation simplicity, security robustness, and computational overhead. Channel-based passive mechanisms are lightweight but vulnerable to environmental changes. CR mechanisms offer stronger security but require real-time updates. Tag-based schemes provide flexibility but demand careful key management. Future research should focus on selecting the appropriate mechanism via specific application requirements.

\begin{figure}[t] 
	\centering  
	\includegraphics[width=0.85\linewidth]{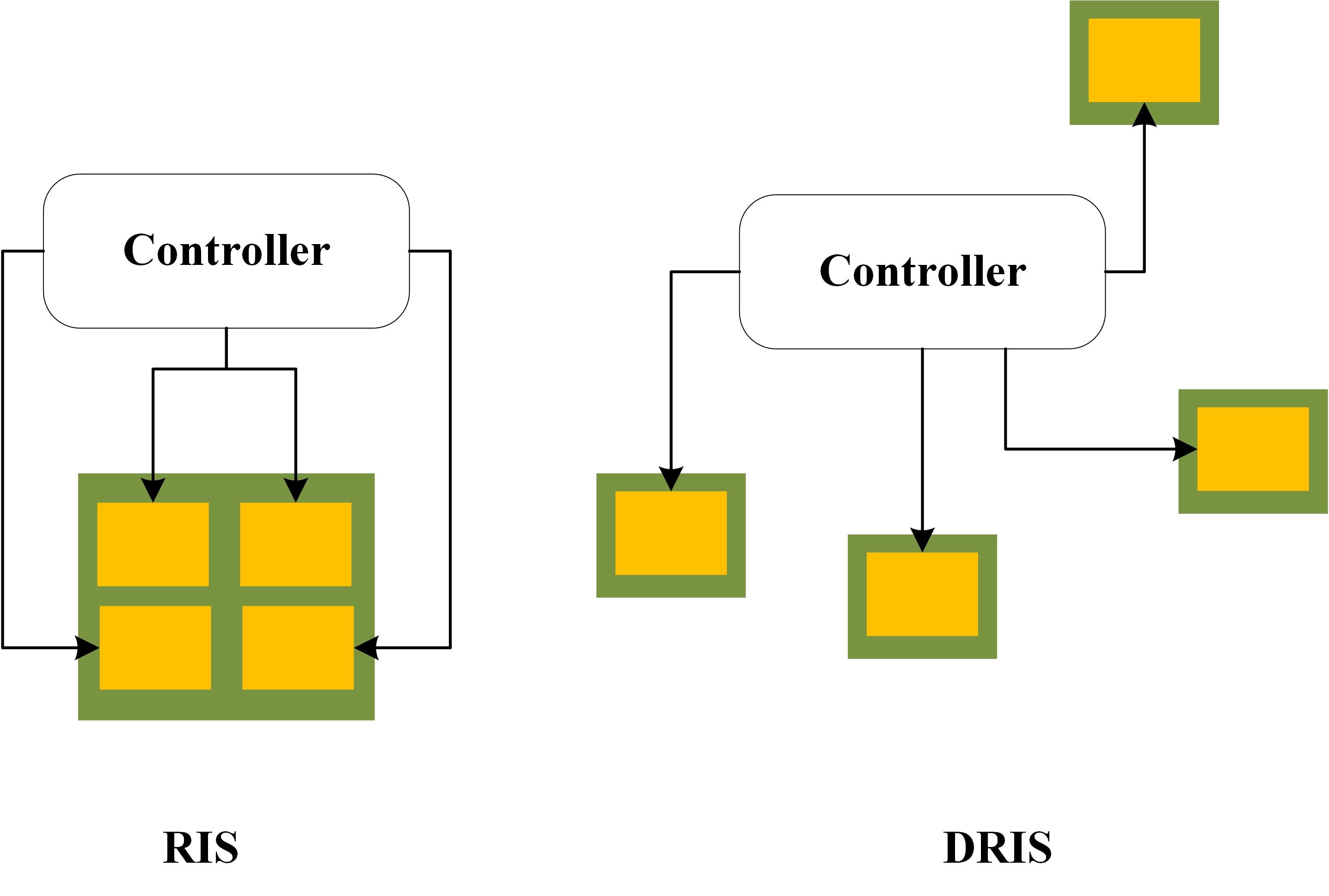} 
	\caption{Pictorial representation of the difference between RIS and DRIS.} %
	\label{DRIS} %
\end{figure}

\subsubsection{RIS-assisted PLA Mechanism Utilizing Other Innovative Architectures}
Table \ref{tbl5} summarizes a comparison of the three RIS-assisted PLA mechanisms. In contrast to the previously discussed RIS-assisted PLA, some studies have begun to investigate architectural innovations and cross-layer integration to enhance the performance and security of RIS-assisted PLA.
\begin{itemize}
	\item \textbf{H-RIS-assisted authentication mechanism:}
	Selim et al. propose a PLA method that leverages the unique signal reflection and sensing capabilities of H-RIS to effectively enhance authentication performance \cite{ref68}. Unlike traditional RIS, which only reflects signals, H-RIS can simultaneously reflect and sense signals \cite{refH-RIS-8, refH-RIS-9, refH-RIS-10}, thereby obtaining richer channel information and improving the distinguishability between legitimate channels and attack channels. This method includes two stages: channel identification and channel verification. In the channel identification stage, both H-RIS and the receiver Bob simultaneously perform channel estimation, using H-RIS's sensing capabilities to obtain detailed channel characteristics for accurate identification of Alice's channel. In the channel verification stage, the generalized likelihood ratio test (GLRT) is used to compare the reference channel with the current channel. The additional information provided by H-RIS significantly enhances the ability to distinguish between Alice and the attacker Eve, reducing the probability of misjudgment.
	\item \textbf{Distributed RIS-assisted authentication mechanism:}
	The PLA scheme proposed by Brighente et al. employs Distributed Reconfigurable Intelligent Surface (DRIS) technology to enhance the authentication performance of Visible Light Communication (VLC) systems \cite{refVLC,refDRIS}. The architecture of the DRIS is illustrated in Fig. \ref{DRIS}. In contrast to traditional RIS, DRIS features a wider distribution of reflecting elements, thereby enhancing the spatial diversity of the VLC channel model and overcoming its limitations. This design expands the scope of channel information acquisition and enables legitimate transmitters to send challenge signals to the receiver through time allocation. DRIS provides richer spatial features in low SNR environments, making it difficult for attackers to replicate challenge signals.
	\item \textbf{RIS-assisted authentication mechanism with cross-layer design:}
	Shawky et al. propose a RIS-assisted PLA scheme incorporating cross-layer design to enhance authentication performance in vehicular communication \cite{refcross} . Unlike traditional authentication methods that rely solely on physical-layer channel characteristics, this scheme integrates Public Key Infrastructure-based initial legitimacy detection and key-based physical-layer re-authentication, thereby constructing a cross-layer authentication framework \cite{refcross-26}. Within this framework, the RIS is utilized to enhance the SNR for targeted vehicles, thereby improving PD of PLA, especially in Non-line-of-sight (NLoS) scenarios. Through theoretical analysis and experimental validation, the results demonstrate that the RIS can significantly improve detection probability at low SNR. Furthermore, Burrows-Abadi-Needham (BAN) logic is used to demonstrate that the scheme effectively resists passive and active attacks.
	\item \textbf{RIS-assisted device feature PLA:}
	Illi et al. propose a robust RIS-based PLA scheme that leverages Carrier Frequency Offset (CFO) as a hardware characteristic for authentication \cite{refR20}. The scheme optimizes RIS phase configuration to enhance signal reception for legitimate users while mitigating eavesdropping risks under confidentiality constraints \cite{refR20-27}. The findings show that increasing the size of the RIS array and the CFO difference between authorized and unauthorized transmitters strengthens authentication robustness.
\end{itemize}

\subsection{Potential Security Threats of RIS-assisted PLA}
The four aforementioned RIS-assisted PLA mechanisms employ distinct authentication principles and design approaches, yet they remain susceptible to fundamental security threats inherent to RIS-enabled systems. This subsection systematically identifies and discusses two foundational security threats that can compromise any of the above authentication mechanisms, namely RISJ and RISL.

\subsubsection{RISJ attack}
The RIS jamming (RISJ) attack is an active threat in which an attacker interferes with the proper functioning of a communication channel by \emph{unauthorizedly} manipulating the RIS reflection matrix \cite{refRISJ1}. As a result, the receiver cannot accurately measure channel characteristics consistent with the expected RIS configuration, thereby disrupting authentication of the legitimate sender's identity. For instance, an attacker may induce a reflection pattern that deviates from the legitimate configuration, directly interfering with the PLA process. The primary consequences include frequent authentication failures and false positives; in extreme cases, it may even prevent legitimate users from accessing the system. Furthermore, such disruptions enable attackers to masquerade as legitimate users, triggering identity obfuscation. We summarize the defense strategies as follows:
\begin{itemize}
	\item {Channel separation technique: Leverages the disparity between the direct channel and the RIS-reflected (cascaded) channel under \emph{expected} RIS configurations to identify abnormal inconsistencies \cite{refRISJ2}.}
	\item {Multi-path feature monitoring: Enhances the system's ability to discern malicious RIS interference by assessing channel characteristics across diverse subcarriers, spatial dimensions, or temporal coherence, while remaining robust against legitimate configuration updates.}
\end{itemize}

\subsubsection{RISL attack}
A RIS leakage (RISL) attack is a passive threat in which an attacker obtains, leaks, or steals channel information associated with a legitimate communication link by controlling the RIS \cite{refRISL1}. By learning channel-related information that can reproduce or predict physical-layer signatures, the attacker can masquerade as a legitimate user, compromising authentication integrity. Typically, RISL is realized through learning-driven forgery: the attacker repeatedly probes the environment via RIS control, observes the resulting receptions, and infers the relationship between RIS-induced reflections and cascaded channel effects. Based on these observations, the attacker generates synthetic signature data and constructs false ``channel fingerprints'' to deceive the authentication system, even without knowing the instantaneous RIS configuration. We summarize the defense strategies as follows:
\begin{itemize}
	\item {Dynamic private pilot design: Adjusts the time and frequency characteristics of pilot signals to prevent attackers from obtaining consistent channel measurements through a RIS \cite{refRISL2}.}
	\item {Cooperative RIS framework: Deploys multiple trusted RISs and dynamically coordinates them to constrain the attacker's ability to learn a stable mapping between RIS control and channel fingerprints.}
\end{itemize}

Table~\ref{tbl6} summarizes the key differences between RISJ and RISL attacks in terms of attack type, mechanism, effects, and defense strategies.

\begin{table*}[t]
	\caption{Comparison of RISJ and RISL Attacks on PLA\label{tbl6}}
	\centering
	\renewcommand{\arraystretch}{1.5}
	\begin{tabular}{|m{3.3cm}|m{0.39\textwidth}|m{0.36\textwidth}|}
		\hline
		\textbf{Aspect} & \textbf{RISJ} \cite{refRISJ1,refRISJ2} & \textbf{RISL} \cite{refRISL1,refRISL2} \\
		\hline
		Attack Type & Active: Disrupts communication by unauthorized manipulation of RIS paths & Passive: Steals/learns channel information via RIS control \\
		\hline
		Mechanism & Induces reflection-path/configuration inconsistency, causing authentication errors & Enables forgery by learning/inferencing channel fingerprints from observed RIS-induced effects \\
		\hline
		Key Effects & $\bullet$ Authentication failures. \newline $\bullet$ False positives and impersonation risks & $\bullet$ High-precision channel knowledge extraction \newline $\bullet$ Identity forgery \\
		\hline
		Defense Strategies & $\bullet$ Channel separation under expected RIS configurations \newline $\bullet$ Multipath feature monitoring with temporal/spatial consistency checks & $\bullet$ Dynamic private pilot design \newline $\bullet$ Cooperative RISs frameworks to limit learnability and prediction \\
		\hline
	\end{tabular}
\end{table*}

\subsection{Summary}
We investigate RIS hardware architectures, authentication mechanisms, and security challenges and observe that the primary gain of RIS-assisted PLA does not stem from conventional transmit-power improvements or a single-point enhancement in receiver signal processing. The programmable capability of reconfigurable intelligent surfaces to reshape the propagation environment is the main source of benefit. By increasing the controllability and complexity of channel features, the distinguishability between legitimate links and potential attackers can be improved, thereby enhancing authentication reliability. However, we also observe that such programmability introduces new risks of manipulation, which necessitate more robust verification mechanisms to better resist adversarial actions.

\section{Spatial-Domain Summary and Future Research Directions}
\label{sec6}
In this section, we provide a concise spatial-domain overview to highlight how different spatial-domain mechanisms can enrich PLA features and explore several emerging technologies with significant potential to further advance PLA. We examine three promising research directions, including metamaterials-assisted PLA, XL-MIMO-assisted PLA, and FAS-assisted PLA. For each future direction, we analyze the underlying principles, present representative schemes, and discuss their unique contributions to enhancing authentication security, reliability, and applicability in next-generation wireless systems.

\subsection{Concise Comparison of Spatial-Domain Technologies}

\begin{table*}[t]
	\caption{Comparative Analysis of Spatial Enhancement Technologies Central to This Survey\label{tab:space_tech_comparison}}
	\renewcommand{\arraystretch}{1.25} 
	\centering
	\begin{tabular}{|m{0.15\textwidth}|m{0.24\textwidth}|m{0.26\textwidth}|m{0.26\textwidth}|}
		\hline
		\textbf{Category} & \textbf{DPA} & \textbf{Massive MIMO} & \textbf{RIS} \\
		\hline
		\textbf{Technique} & DPA & Massive MIMO & RIS \\
		\hline
		\textbf{Operating Domain} & Space + polarization & Spatial Array & Spatial Environment \\
		\hline
		\textbf{Control Position} &
		Device-side\newline(Antenna hardware/architecture) &
		Device-side\newline(Array-based sensing/processing) &
		Environment-side\newline(External control) \\
		\hline
		\textbf{Physical Mechanism} &
		$\bullet$ Transmit/receive using orthogonal polarization states\newline
		$\bullet$ Polarization-dependent channel response shaped by depolarization effects &
		$\bullet$ Large-scale antenna array sensing and signal processing\newline
		$\bullet$ Enables fine-grained spatial/statistical observations (e.g., channel hardening and multi-antenna statistics) &
		$\bullet$ Programmable propagation control via reconfigurable reflecting elements\newline
		$\bullet$ Reshapes propagation paths and cascaded channel realizations (possibly under random configuration) \\
		\hline
		\textbf{Enhancement for PLA} &
		$\bullet$ Introduces an additional polarization feature dimension for authentication\newline
		$\bullet$ Provides polarization-dependent signatures\newline
		$\bullet$ Improves discrimination under low-SNR and fine channel variations &
		$\bullet$ Increases spatial resolution and the richness of observable channel statistics\newline
		$\bullet$ Enhances uniqueness via fine-grained multipath/hardware-induced features used in statistical hypothesis testing\newline
		$\bullet$ Supports authentication without relying on attacker CSI &
		$\bullet$ Adds controllable spatial DoF to modify the channel/feature realizations used by PLA\newline
		$\bullet$ Improves authentication reliability by enabling challenge-response style verification and/or stronger channel-feature separability through RIS-induced signatures\newline
		$\bullet$ Extends PLA to scenarios with weak/blocked direct links by leveraging cascaded/direct channel features \\
		\hline
	\end{tabular}
\end{table*}

As summarized in Table~\ref{tab:space_tech_comparison}, the key distinction among these three spatial-domain technologies lies in where the control is applied and what physical quantity is enriched. 
DPA enriches the authentication feature space at the device side by introducing a polarization-dependent channel response in addition to spatial fading. Massive MIMO further enriches feature richness at the device side through large-scale array-based sensing and processing, enabling fine-grained spatial/statistical observations that improve statistical distinguishability in hypothesis testing.
Finally, RIS performs programmable propagation control at the environment side by reconfigurably reshaping propagation paths and cascaded channel realizations, thereby providing additional spatial degrees of freedom and enabling RIS-induced signatures for more robust PLA.

\subsection{Metamaterials-assisted PLA}

Metamaterial technology enables dynamic control of the electromagnetic properties of the beam, allowing the electromagnetic environment to be customized and reconfigured for channel control.

\begin{figure}[t]
	\centering  
	\includegraphics[width=0.85\linewidth]{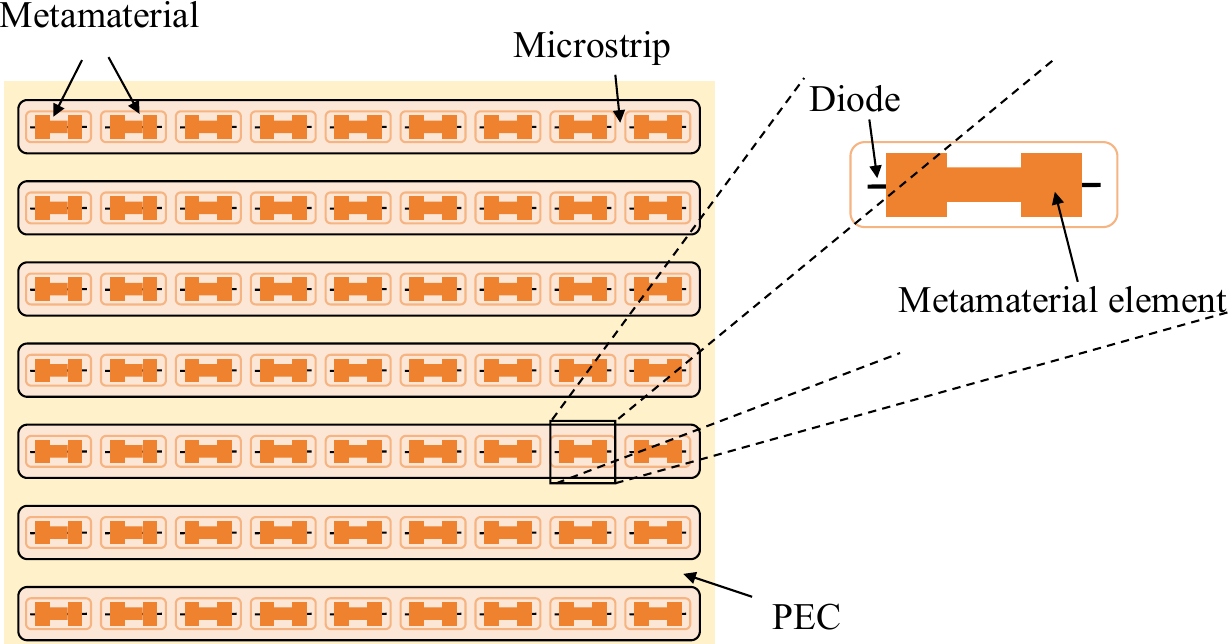}
	\caption{The structure of the DMA.}
	\label{DMA_structure}
\end{figure} 

\subsubsection{DMA-assisted PLA}

DMA is a large-scale antenna array composed of metamaterial elements, as shown in Fig.~\ref{DMA_structure}. DMA utilizes advanced analog signal processing capabilities to programmatically regulate received and transmitted signal beams, so that it achieves excellent performance in large-scale antenna arrays at low cost and power overhead~\cite{refDMA17,refDMA18,refDMA19,refDMA20,refDMA21}. Compared with RIS, DMA offers advanced signal transceivers and processing capabilities that facilitate precise channel estimation~\cite{refDMA9} and rapid beam regulation~\cite{refDMA22}.

Recent studies have demonstrated the advantages of DMA in PLA. For instance, Wang et al. proposed a multi-path PLA scheme based on DMA, employing atomic norm minimization and peak spectral search algorithms to estimate the multi-path information of the transmitted signal~\cite{refDMAPLA1}. Independent authentication is performed for each multi-path channel to optimize information utilization, thereby improving authentication probability. Simulation results indicate that the proposed method outperforms existing multi-path combining authentication schemes, which demonstrates the potential of dynamic metasurface antennas to enhance security.

\begin{figure*}[t]  
	\centering  
	\includegraphics[width=0.8\textwidth]{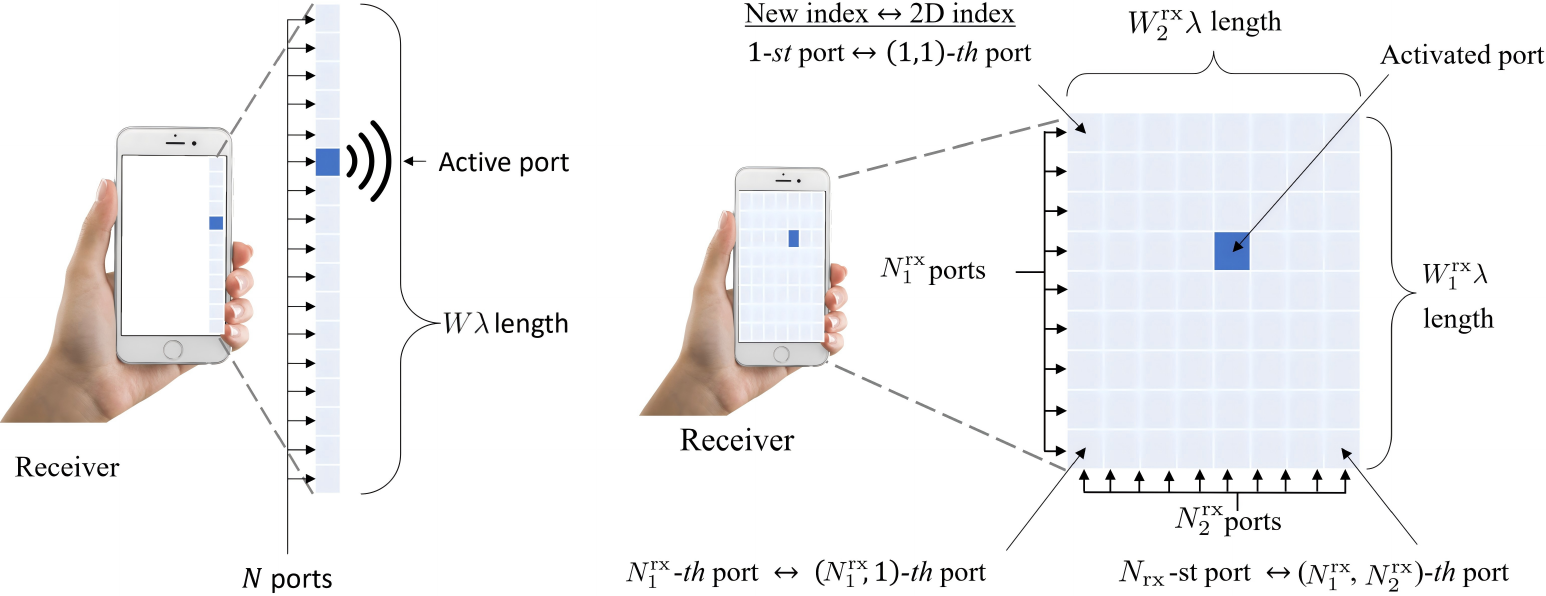}
	\caption{A schematic of both 1D and 2D fluid antenna structures~\cite{FAS-1}.}
	\label{FAS_1_2}
\end{figure*}  
\subsubsection{Metasurface RFF injection (MeRFFI)-assisted PLA}
Conventional RFF technology faces challenges including non-configurability, low reliability~\cite{refMeRFFI20}, limited recognition capability~\cite{refMeRFFI21}, and complex data processing~\cite{refMeRFFI21,refMeRFFI22}. To address these, Rajendran et al. proposed MeRFFI (Metasurface RF-Fingerprinting Injection)~\cite{refMeRFFI}, which utilizes the hypersurface technique~\cite{refMeRFFI8} to inject RF fingerprints at the wireless physical layer, enhancing the security of static IoT devices with low cost (below \$0.2) and low power consumption (below 50 mJ).

MeRFFI enables reliable authentication, ushering in a new paradigm for physical-layer security. Optimizing the metasurface design is anticipated to facilitate its integration into various IoT applications.

\subsection{XL-MIMO-assisted PLA}
XL-MIMO overcomes the stationary channel assumption of conventional massive MIMO by extending the antenna array length, exhibiting pronounced channel non-stationarity and enriched spatial distinguishability in near-field environments~\cite{M10-2,M10-3}.

To address these non-stationary characteristics, Demirci et al.~\cite{M10} introduced the concept of Visibility Region (VR)~\cite{M10-3} into PLA and proposed a VR-based authentication method. By extracting features from the power domain and beam domain, this approach characterizes VR boundaries, spatial extent, and overlap with other users. The spatial extension feature measures the ratio of the VR birth point to its death point, while the discriminative interaction feature quantifies the overlap ratio of each VR. These features are combined into a feature vector for each VR, and the final dynamic VR signature vector is constructed by sorting all VR vectors by their non-overlapping lengths.

During authentication, the base station compares the current VR signature with the registered template and employs a binary decision process to detect impersonation or attacks. Simulation results show that the proposed method improves attack detection probability by 87\% and 70\% in static and dynamic channel scenarios, respectively, significantly outperforming traditional schemes~\cite{refM5, M10-6}. Consequently, XL-MIMO demonstrates tremendous potential in spectral efficiency, spatial division multiple access~\cite{M10-2,M10-11}, and physical layer security.

\subsection{FAS-assisted PLA}
\subsubsection{Unique space-time fingerprint}
The FAS enables continuous, rapid, and controllable spatial reconfiguration of the radiating port by sliding liquid metal within a micro-scale conduit~\cite{FAS-1}. As shown on the left side of Fig.~\ref{FAS_1_2}, this implements a 1D linearly reconfigurable structure. This dynamic spatial reconfigurability introduces unprecedented temporal variability and unpredictability to the communication channel. As the FAS port switches positions in space and time, the receiver acquires a set of highly random channel responses during each authentication cycle. At the $t$-th authentication slot, Bob receives the signal through port $k_t$ over the corresponding channel coefficient $h_{AB,k_t}$.

The time-varying channel sequence $\{h_{AB,k_1},\dots,h_{AB,k_T}\}$ constitutes the unique space-time fingerprint of the Alice--Bob link. This fingerprint is accurately reproduced only when Alice is within the nominal region and the spatial topology remains invariant. Port-to-port spatial correlation, modeled by the zeroth-order Bessel function, decays with increasing Euclidean distance between ports, further enhancing fingerprint distinctiveness.

\subsubsection{CR authentication protocol}
We adopt a CR framework with Bob as the verifier and Alice as the prover. The protocol proceeds as follows:
\begin{enumerate}[label=\textcircled{\small\arabic*}]
	\item Bob generates a random port-switching sequence $\mathcal{S}=\{k_1,\dots,k_T\}$ of length $T$ and sends it to Alice.
	\item Alice transmits the pilot symbols $x(t)$ in the prescribed order.
	\item Bob measures $\hat{h}_{AB,k_t}$ at port $k_t$ and reconstructs the channel fingerprint.
	\item Bob computes the decision metric
	\begin{equation}
		D_{\mathrm{auth}} = \frac{1}{T} \sum_{t=1}^T \left| \hat{h}_{AB, k_t} - h_{AB, k_t}^{\mathrm{ref}} \right|^2 
		\mathop{\gtrless}_{H_1}^{H_0} \gamma,
	\end{equation}
	where $\hat{h}_{AB,k_t}$ is Bob's estimate in this epoch, $h_{AB,k_t}^{\mathrm{ref}}$ is the stored reference fingerprint, and $\gamma$ is the decision threshold.
	\item Accept Alice if $D_{\mathrm{auth}}\le\gamma$; otherwise reject.
\end{enumerate}

Bob may optimize $\mathcal{S}$ to maximize the statistical divergence between genuine and adversarial fingerprints, enhancing detection performance.

\subsubsection{Synergy with other spatial technologies}
When combined with RIS or DMA, FAS enables dual-sided spatial randomization. The composite channel at each slot includes both the direct Alice--Bob path and the RIS-cascaded path shaped by the time-varying RIS phase matrix and the FAS port index. Within this full-space reconfiguration framework, communication parties can jointly design port scheduling and environmental programming policies, significantly increasing the authentication security threshold and system adaptability to complex environments.

\subsection{AI-optimized Spatial Configuration}
The unified theoretical framework established in Section~\ref{sec2} introduces the system configuration parameters $\varOmega$, which directly shape the received signal and thereby govern the separability between legitimate and adversarial features. As analyzed above, emerging spatial-domain technologies introduce a high-dimensional and dynamically controllable $\varOmega$ space. As its dimensionality grows with system scale, exhaustive search or hand-crafted configurations become computationally prohibitive, motivating AI methods to directly optimize $\varOmega$ for maximum separability. Two representative approaches utilizing AI are presented as follows:

\subsubsection{Configuration using DRL}
Deep reinforcement learning (DRL) provides a natural framework for addressing this dynamic $\varOmega$ configuration problem~\cite{refYuEdgeIoT2021}, which can be modeled as a Markov decision process. The state captures the current environment and attack conditions, the action specifies the configuration for the next authentication epoch, and the reward encodes the feature-separation objective. A policy network learns to map environmental states to near-optimal $\varOmega$ in an online manner.

\subsubsection{Configuration using generative AI}
Beyond sequential decision-making, generative AI models serve as another effective tool for synthesizing $\varOmega$~\cite{refGenAI}. For instance, a conditional diffusion model can learn the distribution of effective configurations given the environment and a target separability, which enables the sampling of novel $\varOmega$ candidates without requiring online search.

\section{Conclusion}
\label{sec7}
This survey provides a systematic examination of how spatial domain technologies such as DPA, massive MIMO, and RIS reshape feature construction in PLA. Rather than treating these technologies as isolated solutions, we analyze their specific operational mechanisms. Specifically, DPA introduces polarization-domain features independent of spatial information; massive MIMO enhances feature richness by leveraging device-level spatial observations; and RIS modifies authentication features via environment-side programmability. Each mechanism provides specific authentication advantages while introducing distinct security risks that require targeted mitigation. This survey catalogs representative authentication methods for each technology, identifies associated threats and defenses, and explores opportunities from emerging technologies, including AI-optimized spatial configuration. To illustrate the practical implications, case study results compare authentication performance across different parameters and operational scenarios, demonstrating concrete benefits of spatial-domain enhancements. By examining the relationships among enhancement mechanisms, feature extraction methods, and introduced vulnerabilities, this survey supports the secure and efficient deployment of spatial-domain enhanced PLA in physical-layer security frameworks.


\vfill

\end{document}